\documentclass[reprint, amsmath,amssymb, aps, pra]{revtex4-1}

\usepackage{epsfig}
\usepackage{graphicx}
\usepackage{dcolumn}
\usepackage{bm}

\usepackage[usenames,dvipsnames]{xcolor}

\begin{document}

\title{Precision determination of isotope shifts in ytterbium and implications for new physics}
 






\author{N. L. Figueroa$^1$, J. C. Berengut$^2$, V. A. Dzuba$^2$, V. V. Flambaum$^2$, D. Budker$^1$, D. Antypas$^{1,}$}
\email{dantypas@uni-mainz.de}
\affiliation{$^1$ Johannes Gutenberg-Universit{\"a}t Mainz, Helmholtz-Institut Mainz, GSI Helmholtzzentrum f{\"u}r Schwerionenforschung, 55128 Mainz, Germany\\$^2$ School of Physics, University of New South Wales, Sydney, New South Wales 2052, Australia}



\date{\today}

\begin{abstract}

We report measurements of isotope shifts for the five spinless Yb isotopes on the \mbox{$6s^2\,^1\textrm{S}_0 \rightarrow 5d6s\,^1\textrm{D}_2$} transition using Doppler-free two-photon spectroscopy.
We combine these data with existing measurements on two transitions in Yb$^+$ [Phys. Rev. Lett. 125, 123002 (2020)], where deviation from King-plot linearity showed hints of a new bosonic force carrier at the 3$\sigma$ level. 
The combined data strongly reduces the significance of the new-physics signal.
We show that the observed nonlinearity in the joint Yb/Yb$^+$ King-plot analysis can be accounted for by the deformation of the Yb nuclei.


\end{abstract}

\pacs{Valid PACS appear here}

\maketitle
\noindent
\textit{Introduction---} Precision measurements in atomic systems offer ways to study particle physics at low energy \cite{RobertsARNPS2015, DeMilleScience2017, SafronovaADP2019} and complement work involving high-energy accelerators, neutrino studies  and astrophysical observations. Indeed, a number of experiments
employing atomic or molecular probes have yielded a range of tests of fundamental principles and searches for physics beyond the Standard Model (SM) (see review \cite{SafronovaRMP2018} and references therein).
The role of precision spectroscopy in such work is further emphasized by recent ideas pointing to the manifestation of several beyond-SM physics scenarios in the isotope shifts (IS) of atomic spectra \cite{FrugiuelePRD2017,BerengutPRL2018}.

The IS in the frequency  of an atomic transition are approximately linearly correlated with IS observed in another transition.  This linearity in the so-called King-plot analysis \cite{KingJOSA1963, King1983} is used, for example, to extract the variation of nuclear charge radii from atomic spectra \cite{Angeli2013}. While deviations from linearity have long been known \cite{GriffithJOSAB1979} in cases where subdominant nuclear effects 
cannot be neglected, the concept of  nonlinearity (NL) has been recently revisited. As pointed out in \cite{BerengutPRL2018}, NL may arise due to beyond-SM interactions, such as the exchange of light force mediators between  the electron and neutron \cite{DelaunayPRD2017}. Such interaction is invariant under spatial inversion (it is parity-even) and can be sensitively probed for a mediator with Compton wavelength greater than the size of the nucleus, i.e. a  mass $\lesssim 100$ MeV/$c^2$ \cite{DelaunayPRD2017, BerengutPRL2018}. Its parity-odd counterpart has been explored with atomic parity violation studies \cite{DzubaPRL2017,AntypasNP2019,AntypasPRA2019}.

Following the ideas in \cite{FrugiuelePRD2017,DelaunayPRD2017}, in Ref.\,\cite{BerengutPRL2018} a framework was established for constraining new electron-neutron interactions using bounds on King-plot NL. The predictive power of the method, however, depends on the size of SM nuclear effects, that may also give rise to NL.
Such effects were studied in   \cite{FlambaumPRA2018}, while in \cite{BerengutPRR2020}, it was shown that the limitations due to SM effects introducing NL can be 
mitigated if IS data from more than two transitions are combined in  a generalized King-plot analysis.

As availability of high-accuracy IS data enables probing for new bosons, several efforts were initiated to this end  
\cite{OrigliaPRA2018, MiyakePRR2019, ManovitzPRL2019,KnollmannPRA2019,CountsPRL2020, SolaroPRL2020,RehbehnPRA2021}, dramatically improving the measurement accuracy previously limited to $\approx$100\,kHz in the IS \cite{GebertPRL2015}. Most notably,  two experiments with Yb$^+$ \cite{CountsPRL2020} and Ca$^+$ \cite{SolaroPRL2020} reported on high-precision IS measurements, each employing five stable nuclear-spin-zero isotopes \footnote{To observe NL, independent IS data on at least three isotope pairs are needed, so at least four isotopes are required. Isotopes with zero nuclear spin are easier for this analysis, as there is no need to separate the IS from hyperfine structure.}. 
With these two experiments demonstrating comparable sensitivity to new physics (NP), it is intriguing that no King-plot NL was found in the Ca$^+$ work, but NL at the 3$\sigma$ level was observed in Yb$^+$. This discrepancy hints at nuclear effects as the source of the NL in Yb$^+$, with the quadratic field shift suggested as a possible cause in  \cite{CountsPRL2020} and the influence of nuclear deformation on  the field shift proposed in \cite{AllehabiPRA2021}.

While modeling nuclear effects is needed to gain insight into the cause of the observed Yb$^+$ NL, further precision IS data may allow separation of the SM and NP effects in a model-independent way \cite{BerengutPRR2020}. Here we report on IS spectroscopy carried out  in the $6s^2\,^1\textrm{S}_0 \rightarrow 5d6s\,^1\textrm{D}_2$ transition of neutral Yb, employing all five spinless isotopes. The data are combined with those of Ref.\,\cite{CountsPRL2020} to provide a joint Yb/Yb$^+$ King-plot analysis \cite{BerengutPRR2020}. Since the two systems have the same nucleus but different electronic structure, this IS comparison offers a way to evaluate modeling of higher-order nuclear effects, since the IS is generally different for the energy levels of the neutral and ionic Yb. 

\begin{figure}[htb]
\includegraphics[width=8.5cm]{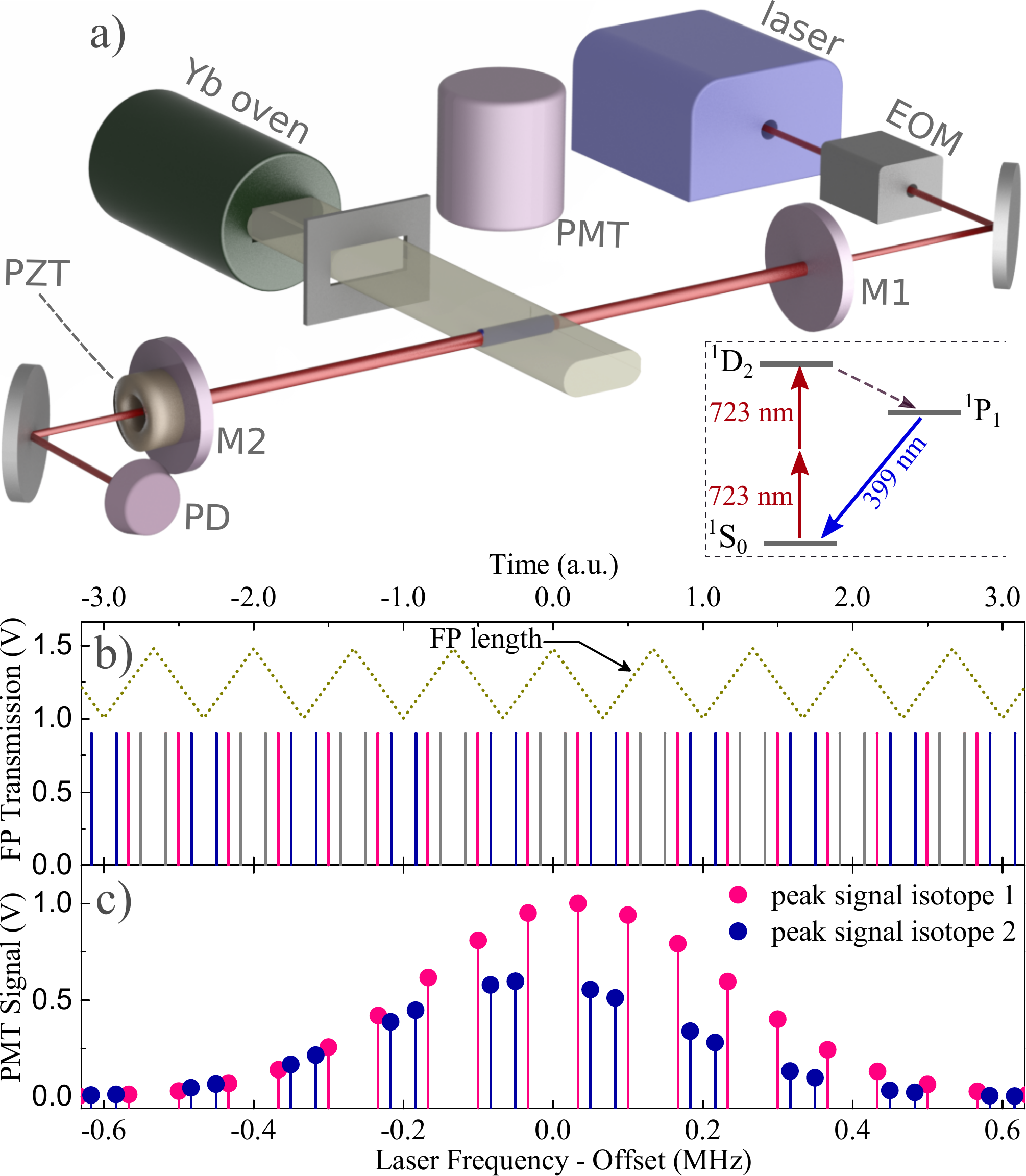}
\caption{\small{
Apparatus and spectrum measurement method. a) Atomic beam setup for intracavity $^1\rm{S}_0\rightarrow \,^1\rm{D}_2$  spectroscopy. EOM: electro-optic modulator; M1, M2: FP cavity mirrors; PMT: photomultiplier; PZT: piezo-electric transducer; PD: photodiode.  b) Simulated FP transmission signal, corresponding to the FP length being rapidly modulated (dotted triangle wave). The carrier (pink) and a  $1^{\rm{st}}$-order sideband (blue) are used to excite a pair of Yb isotopes, while the other  $1^{\rm{st}}$-order sideband (gray) is not resonant with atoms. c) Simulated fluorescence of the atomic beam, resulting from the FP resonances shown in b).  }}
\label{fig:fig1}
\end{figure}

\noindent
\textit{Experiment---} Two-photon, Doppler-free spectroscopy \cite{Demtroder2015Laser2}  is carried out to extract the IS among the five spinless Yb isotopes (with mass number $A$-abundance: 176-12.9\%, 174-31.9\%, 172-21.8\%, 170-3\%, 168-0.1\%). The \mbox{$6s^2\,^1\rm{S}_0\rightarrow\,5d6s\,^1\rm{D}_2$} transition, of $\approx$24\,kHz natural linewidth \cite{BowersPRA1996}, is excited in an atomic beam (Fig.\,\ref{fig:fig1}a) with light at $\lambda\approx$723\,nm and detected via fluorescence at 399\,nm, emitted in the second step of the $5d6s\,^1\rm{D}_2\rightarrow 6s6p\,^1\rm{P}_1\rightarrow\,6s^2\,^1\rm{S}_0$ cascade (Fig.\,1a). Only the component between the $m=0\rightarrow m'=0$ magnetic sublevels is  driven, using linear polarization for the optical field set along a  \mbox{$B\approx 4$\,G} applied magnetic field. 

Our spectroscopy scheme employs two key features, implemented to minimize systematic effects from imperfections in the laser frequency calibration, and due to the $1^{\rm{st}}$-order Doppler effect. First, a pair of isotopes to be measured for their IS are excited concurrently \cite{KnollmannPRA2019}, using phase-modulated light  which is produced with an electro-optic modulator that imposes frequency sidebands at the nominal IS frequency. The resulting optical field has spectral components \mbox{$\nu_{\rm{L}}=\nu_{\rm{C}}\pm n  \nu_{\rm{m}}$}, where $\nu_{\rm{C}}$ is the carrier, $n=0,1,2...$ is the sideband order and $\nu_{\rm{m}}$ is the frequency of the phase modulation, nominally set to match the IS frequency. To excite a pair of isotopes, the carrier and one of the  sidebands with $n=1$ are typically used.
 
The second key feature is that atoms are excited within a scanning Fabry-P\'erot (FP) cavity. Since the counter-propagating components of the intracavity 723-nm field have well-matched wavefronts, the $1^{\rm{st}}$-order Doppler shift is suppressed. The FP has another essential role. It allows to temporally separate excitation of the isotopes being probed and detect the respective fluorescence signals in discrete channels; this enables `simultaneous' probing of both isotopes. Figures~\ref{fig:fig1}b,c illustrate this. As the FP length is rapidly scanned while the laser frequency is slowly swept across the $\rm{S-D}$ resonance, the various spectral components of light become resonant in the FP at different times, thus the two isotopes are probed in a rapid, but resolved time sequence. Monitoring the FP transmission is done to gate detection using the respective FP fringes as gates. The peak signal within a gate is used to construct a lineshape profile for each isotope.
 
To extract the IS, the frequency $\nu_{\rm{m}}^0$ is determined that yields null frequency offset between the two isotope lineshapes [see Supp. Mat.].  In the absence of systematics, the IS for the two-photon transition is $\delta \nu=2\nu_{\rm{m}}^0$. During a measurement run, lineshapes for a given isotope pair are recorded for many values of $\nu_{\rm{m}}$ in a range $\nu_{\rm{m}}^0\pm10$ kHz.  Fitting to the lineshapes is done to determine their frequency offset and its variation with $\nu_{\rm{m}}$, in order to extract $\nu_{\rm{m}}^0$.  The $\pm 2\times$10 kHz variation of the relative lineshape offset is a small fraction of the observed $\approx$800 kHz transition width, that is limited by transit time of atoms through the intracavity field \cite{LehmannJCP2021}. With the lineshapes practically overlapping during the laser sweep, systematics due to nonlinearity of the sweep are avoided [Supp. Mat.]. 
  
Several unwanted effects may introduce systematic shifts and  require careful treatment [Supp. Mat.]. Some are related to the ac-Stark effect. For the typical intracavity power of 5\,W, it causes a shift of order 20 kHz to the lineshapes. With active stabilization of carrier-sideband power ratio, the ratio is kept to unity to within 0.1\%, and the residual mismatch is recorded and included in the analysis. However, parasitic interferences, such as for example between the fundamental transverse cavity mode and high-order modes, can, in principle, result in different intracavity intensity experienced by the two isotopes \cite{AntypasOL2018}. To eliminate potential shifts, data are taken at many powers and extrapolation to zero power is done to extract the IS. 

Another complication arises due to the multiple frequency components in the light injected into the FP. While a single component is resonant in the FP at a given time, one of the off-resonant components entering the FP due to the finite reflectivity of the input mirror ($\approx96.5$\%), acts with the resonant component to cause small, but measurable excitation of the `unwanted' isotope within the detection channel of the  isotope primarily being measured. For example, when the 1$^{\rm{st}}$-order sideband of frequency $\nu_{\rm{L}}=\nu_{\rm{C}}+ \nu_{\rm{m}}$ is resonant and excites an isotope at frequency $2(\nu_{\rm{C}}+ \nu_{\rm{m}})$, due to the off-resonant  sideband with $\nu'_{\rm{L}}=\nu_{\rm{C}}- \nu_{\rm{m}}$, Doppler-free excitation at frequency  $\nu_{\rm{L}}+\nu'_{\rm{L}}=2\nu_{\rm{C}}$  also occurs. 
This unwanted background signal has small impact on extracting the IS between isotopes of similar abundance (e.g. $^{174}$Yb/ $^{172}$Yb), but substantial impact if the abundance varies drastically (e.g. $^{170}$Yb/$^{168}$Yb). In the former case, we correct the IS measurements using Monte-Carlo simulations of the effect; in the latter, we measure the backgrounds and correct the data accordingly.

To check for unaccounted for effects  and ensure consistency of measurements, data are taken under a variety of conditions. For example, with FPs of different lengths, alternating use of the 1$^{\rm{st}}$-order sidebands, and at different Yb oven temperatures. The obtained accuracy in the IS, measured for the four pairs of adjacent mass, is $\approx$300 Hz (710 Hz for the low-abundance pair  $^{170}$Yb$/^{168}$ Yb), 
and is limited by statistical uncertainty. Additional checks using the two 1$^{\rm{st}}$-order sidebands to probe the pairs $^{176}$Yb$-$$^{172}$Yb and $^{172}$Yb$-$$^{168}$Yb are consistent with the main dataset to within 0.3$\sigma$ and 1.4$\sigma$, respectively. 
Our main experimental results are summarised in Table~\ref{tab:MainResults}.

\begin{table}[!tb]
\begin{ruledtabular}
\caption{\label{tab:MainResults}
Isotope shifts of the $^1S_0\rightarrow \, ^1D_2$ transition of Yb. The main results are listed in the first four rows and are used in King-plot analysis, while the remaining are complementary measurements used to check experimental consistency. The shift is defined as $\delta \nu^{AA'}=\nu^{A}-\nu^{A^{\prime}}$.}
\begin{tabular}{ c  c  c  c }
\textrm{Isotope pair (\textit{A-A$^\prime$})}&
\textrm{Measured $\delta \nu^{AA'}$ (kHz)}&
\textrm{Method\footnote{C-S: measurement with use of carrier and $1^{st}$-order sideband fields; S-S: with use of the $1^{st}$-order sidebands.}}
\\
\colrule
168-170 &  1781785.36(71)  & C-S\footnote{Isotope with $A$=168 was always excited by the carrier, in order to mitigate the effects of parasitic backgrounds in the measurements [see Supp. Mat.].} \\
170-172 &  1672021.51(30) & C-S \\
172-174 &  1294454.44(24)  & C-S \\
174-176 &  1233942.19(31)  & C-S\\
\hline
168-172 &  3453805.27(83) & S-S\\
172-176 &  2528396.50(34)  & S-S\\
\end{tabular}
\end{ruledtabular}
\end{table}

\noindent
\textit{King Plots---}
To a good approximation, the IS of a transition $i$ is given by
\begin{equation} \label{eq:basicIS}
\delta\nu^{AA'}_{i} = F_i\delta\langle r^2 \rangle^{AA'} + K_i\mu^{AA'}, 
\end{equation}
where $\delta\langle r^2 \rangle^{AA'}$ is the change in nuclear mean-square charge radius between isotopes $A$ and $A'$, $\mu^{AA'} = 1/m_A - 1/m_A'$ is the change in inverse nuclear mass, and $F_i$ and $K_i$ are the electronic field shift and mass shift constants, respectively. In a King plot~\cite{KingJOSA1963}, modified IS $\delta\nu_i^{AA'}/\mu^{AA'}$ for two transitions are plotted against each other for each independent isotope pair, resulting in a straight line as long as Eq.\,\eqref{eq:basicIS} holds.

\begin{figure}[htb]
\includegraphics[width=\columnwidth]{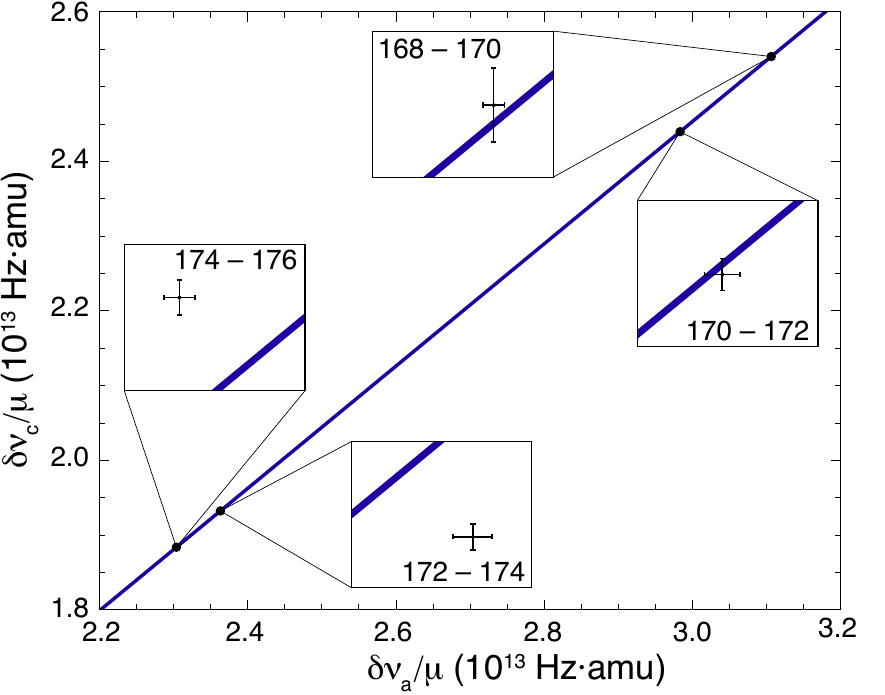}
\caption{King-plot comparison of the measured Yb $^1\textrm{S}_0\rightarrow \, ^1\textrm{D}_2$ transition ($c$) and the Yb$^+$ $^2\textrm{S}_{1/2}\rightarrow \, ^2\textrm{D}_{5/2}$ transition ($a$) \cite{CountsPRL2020}.
The inset graphs have width $5\times10^7$ Hz$\cdot$amu, and show 1$\sigma$ error bars that do not include nuclear mass uncertainty since this uncertainty lies parallel to the slope of the King plot $F_c/F_a$~\cite{CountsPRL2020,SolaroPRL2020}.
The observed deviation of data points from the best-fit line, i.e. the nonlinearity, has $8.2\sigma$ significance.
}
\label{fig:KPlot}
\end{figure}

We have three transitions measured with high precision: $^2\textrm{S}_{1/2}\rightarrow \, ^2\textrm{D}_{5/2}$ and $^2\textrm{S}_{1/2}\rightarrow \, ^2\textrm{D}_{3/2}$ in Yb$^+$ (lines $a$ and $b$, respectively, reported in~\cite{CountsPRL2020}), and $^1S_0\rightarrow \, ^1D_2$ in Yb (line $c$, this work).
In Fig.~\ref{fig:KPlot}, we present a King plot of line $a$ against line $c$ for our four isotope pairs. The line of best fit is obtained by minimizing $\chi^2$ defined as the sum of the error-weighted distance between each point and the line. The best fit corresponds to $\chi^2 = 70.7$ for two degrees of freedom, implying evidence for nonlinearity at the $8.2\sigma$ level.

There are many potential SM sources of NL, as well as NP possibilities. These manifest as additional terms in Eq.~\eqref{eq:basicIS}, so that:
\begin{equation}
\label{eq:IS}
\delta\nu^{AA'}_{i} = F_i\delta\langle r^2 \rangle^{AA'} + K_i\mu^{AA'} + G_i\lambda^{AA'} + \frac{\alpha_\textrm{NP}}{\alpha} D_i \gamma^{AA'}.
\end{equation}
Here $G_i$ represents additional electronic structure factors associated with SM contributions. In this work we consider two higher-order field shift contributions, $G^{(2)}$ and $G^{(4)}$, associated with quadratic field shift ($\lambda = \delta \langle r^2 \rangle^2$) and  nuclear deformation ($\lambda = \delta \langle r^4 \rangle$), respectively.
The quantity $\alpha_\textrm{NP}/\alpha$ ($\alpha\approx 1/137$ is the fine-structure constant) represents the strength of an additional neutron-electron coupling mediated by the exchange of a  light boson of mass $m_\phi$, and $\gamma^{AA'} = A' - A$ is the change in neutron number. The electronic factor $D_i$ must be calculated [Supp. Mat.].

To help distinguish between potential NL sources, and allow us to compare with other transitions in Yb/Yb$^+$, we use the ``frequency-normalised'' King plot and nonlinearity measures $\zeta_{\pm}$ introduced in~\cite{CountsPRL2020}. In this procedure the isotope-pair-dependent quantities in Eq.~\eqref{eq:IS} are normalised to the reference transition $a$: $\overline{x}^{AA'} \equiv x^{AA'}/\nu_a^{AA'}$ for $x \in \{ \delta\nu_c, \mu, \delta \langle r^2 \rangle^2, \delta  \langle r^4 \rangle, \gamma \}$, leading to the linear relationship
\begin{equation}
\label{eq:freqKP}
\overline{\delta\nu}_c = F_{ca} + K_{ca} \overline{\mu} + G^{(2)}_{ca} \overline{[\delta \langle r^2 \rangle^2]} + G^{(4)}_{ca} \overline{\delta \langle r^4 \rangle} + \frac{\alpha_\textrm{NP}}{\alpha} D_{ca} \overline{\gamma},
\end{equation}
where the electronic factors are $F_{ca} = F_c/F_a$ and to lowest order $P_{ca} = P_c-F_{ca}P_a$ for $P \in \{ K, G^{(2)}, G^{(4)}, D\}$. The frequency-normalised King plot is constructed by plotting $\overline{\delta\nu}_c = \delta\nu_c/\delta\nu_a$ against $\overline{\mu}$. This plot has a very small slope, so mass uncertainties along the horizontal axis $\overline{\mu}$ have negligible effect. 

\begin{figure}[tb]
	\includegraphics[width=\columnwidth]{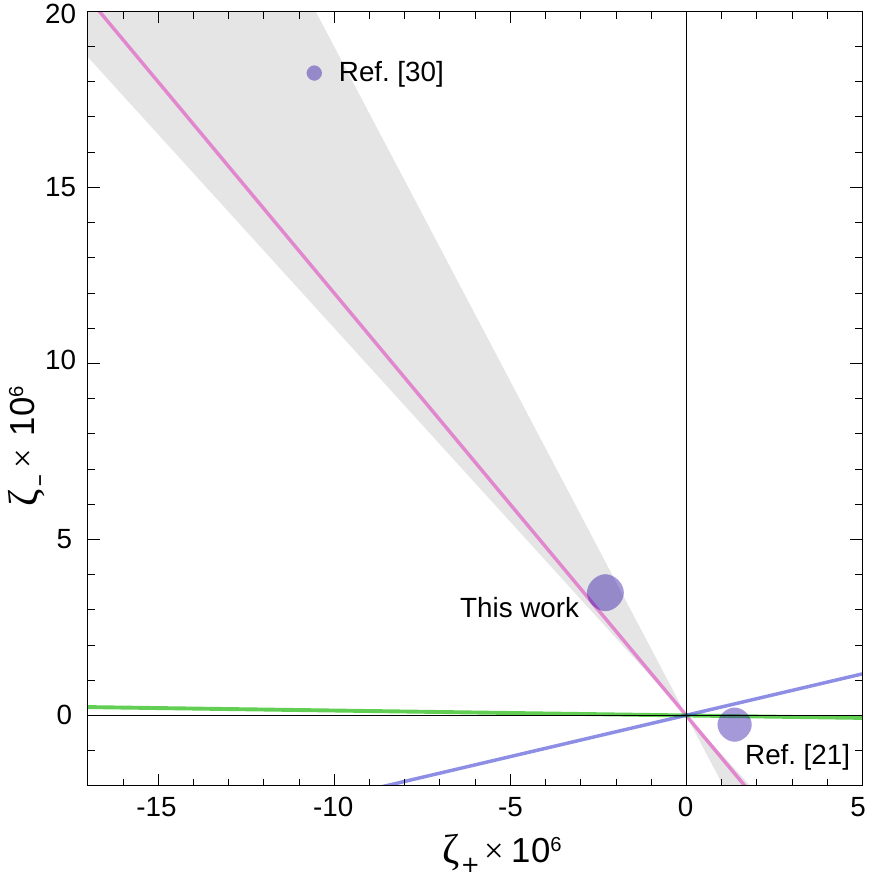}
	\caption{Nonlinearity measures $\zeta_-$ vs $\zeta_+$ ($1\sigma$ range) for next-neighbour isotope pairs. Straight lines show the directions expected from different NL sources --- green line: new physics; blue line: quadratic field shift
	; pink line: nuclear deformation. The shaded region indicates the 1$\sigma$ confidence level for the ratio $\zeta_-/\zeta_+$ from this work. }
	\label{fig:zetaplot}
\end{figure}

\nocite{ono21arxiv}
Vertical deviations of points in the frequency-normalised King plot from the line of best-fit $d_A$ are characterised, as in Ref.~\cite{CountsPRL2020}, using the nonlinearity measures
$$\zeta_{\pm} = d_{168}-d_{170} \pm (d_{172}- d_{174})$$ 
where the subscript $A$ refers to the isotope pair $(A, A+2)$. In Fig.~\ref{fig:zetaplot} we plot $\zeta_-$ against $\zeta_+$ for different transitions (in all cases we use transition $a$, Yb$^+$ $6s \rightarrow 5d_{5/2}$,  as the reference). Also shown are directions in the $\zeta_+ - \zeta_-$ space expected from different NL sources. 
The current data does not support the majority of NL being due to new physics or quadratic field shift, however the possibility of nuclear deformation as the main cause of NL, as discussed in~\cite{AllehabiPRA2021}, is still open. 
To determine the ratio suggested by nuclear deformation, we use experimental values of the nuclear parameters  $\langle r^2 \rangle$ from \cite{Angeli2013} and calculate $\langle r^4 \rangle$ using nuclear deformation parameters $\beta$ \cite{raman01adndt}. We find that $\overline{\delta\langle r^4 \rangle}$ has a ratio of $\arctan(\zeta_-/\zeta_+) = -0.9\,(0.5)$, consistent with our experimental value of $-0.99\,(0.13)$.
We discuss this further in the Supplementary Material.

\noindent
\textit{Generalised King Analysis---}
To place limits on NP we employ the generalised King analysis (GK)~\cite{BerengutPRR2020}. This data-driven method combines the three high-precision transitions in Yb and Yb$^+$ with calculation of $D_i$ factors [Supp. Mat.]. In the absence of NL, the modified IS $\vec{\nu}_{\rm i}\equiv \delta\nu_i^{AA'}/\mu^{AA'}$ for each transition form vectors of dimension 4 (the number of isotope pairs $AA'$) that all lie in a plane along with the unit vector in isotope space, $\vec{\nu}_{\rm \mu} = (1, 1, 1, 1)$. Following~\cite{BerengutPRR2020}, we say the data set is planar provided that the volume $V = \det(\vec{\nu}_{\rm a}, \vec{\nu}_{\rm b}, \vec{\nu}_{\rm c}, \vec{\nu}_{\rm \mu})$ is consistent with zero within experimental errors. The NP coupling is then given by
\begin{equation}
\label{eq:alpha_NP}
\alpha_\textrm{NP} = \frac{V}{V_\textrm{th} (\alpha_\textrm{NP} = 1)},
\end{equation}
where $V_\textrm{th}$ is the volume formed using the theory ansatz \eqref{eq:IS} (see Eq. (13) in Ref.~\cite{BerengutPRR2020}). 
The benefit of the GK analysis is that it accounts for the dominant SM contribution to NL without relying on knowledge of its size. That is, Eq.~\eqref{eq:alpha_NP} does not require calculation of the $G$ parameters or $\lambda^{AA'}$.

\begin{figure}[tb]
\includegraphics[width=\columnwidth]{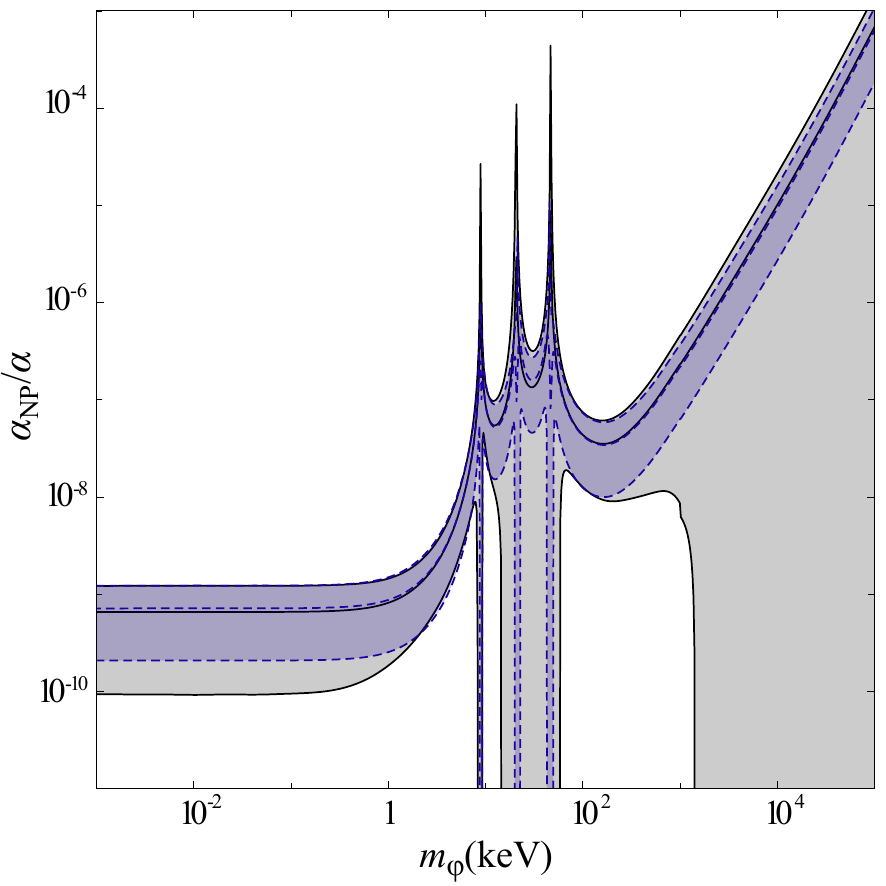}
\caption{
95\% confidence limits  on NP from IS bounds in Yb/Yb$^+$.
Blue-shadowed area:  range of $\alpha/\alpha_{\rm NP}$ (value and bounds indicated by dashed blue lines) from analysis of two Yb$^+$ lines \cite{CountsPRL2020}, where there is a $3\sigma$ NP signal. Grey-shadowed area: range of $\alpha/\alpha_{\rm NP}$ (value and bounds indicated by black lines)  from generalized King-plot analysis of  all three lines, that reduces the significance of NP to $2.3\sigma$ at small $m_\phi$ and $1.8\sigma$ at large $m_\phi$. The reduced significance of NP decreases the lower limit of the allowed NP region. The absence of lower bound indicates that vanishing $\alpha_{\rm NP}$ is consistent with experiment below $2\sigma$.
}
\label{fig:alphaNP}
\end{figure}

In Fig.\,\ref{fig:alphaNP} we show the NP coupling $\alpha_\textrm{NP}$ resulting from this analysis. (To avoid possible confusion with units we present the ratio $\alpha_\textrm{NP}/\alpha$.) The previous results comparing only transitions $a$ and $b$ have a $3\sigma$ NL, while the additional use of transition $c$ decreases the NP ``detection'' to below $2.3\sigma$. This is despite the Yb line showing a large NL when compared in a usual 2D King plot with either of the two Yb$^+$ transitions. The resonance-like structures are points in $m_\phi$ where, due to cancellation of electronic factors $D_i$ from Eq. \eqref{eq:IS}, there is no sensitivity to $\alpha_\textrm{NP}$ and no limits on the allowed range of $\alpha_\textrm{NP}$ can be made. At high mass, the significance of NP is $1.8\sigma$, and so the lower bound in Fig.\,\ref{fig:alphaNP} disappears.

\noindent
\textit{Conclusions---} Our measurement of the $6s^2\,^1\textrm{S}_0 \rightarrow 5d6s\,^1\textrm{D}_2$ transition in Yb shows a significant King plot nonlinearity when compared with the Yb$^+$ transitions of Ref.~\cite{CountsPRL2020}. We have shown that, unlike in~\cite{CountsPRL2020}, the majority of this NL cannot be explained by new physics or quadratic field shift, however it may still be accounted for by nuclear deformation~\cite{AllehabiPRA2021}. Employing the GK analysis to remove the leading SM source of NL reduces the signal for NP to $\sim 2\sigma$.

It is worth pointing out that the Yb transition studied in this work has a rather different electronic structure to the transitions in Yb$^+$ compared in~\cite{CountsPRL2020}. This leads to larger differences in the electronic structure factors and hence higher sensitivity to new physics and SM sources that break King linearity. Thus with similar measurement accuracy to \cite{CountsPRL2020}, we obtain a more significant NL and a higher sensitivity to distinguishing NL sources as shown in Fig.~\ref{fig:zetaplot}.

Future availability of precision data on additional transitions in either the neutral or ionic Yb, may allow for extension of the GK analysis and evaluation of the remaining NL effect. In the final stages of preparing this work, an isotope shift measurement of the Yb $6s^2\,^1\textrm{S}_0 \rightarrow 6s6p\,^3\textrm{P}_0$ was reported~\cite{ono21arxiv}. We have included this data in Fig.~\ref{fig:zetaplot} and note that the nonlinearity ratio $\zeta_-/\zeta_+$ of this transition is consistent with that of the $6s^2\,^1\textrm{S}_0 \rightarrow 5d6s\,^1\textrm{D}_2$ transition in Yb. This suggests a common origin for the two nonlinearities, but does not assist us in determining the cause. Stronger constraints on new physics can likely be obtained if the effect of nuclear deformation is either accounted for or absent (e.g., in isotopes with spherical nuclei).


We gratefully acknowledge V. Vuletic for insightful communications. This project has received funding from the European Research Council (ERC) under the European Union’s Horizon 2020 research and innovation programme (grant agreement No 947696) and the Deutsche Forschungsgemeinschaft (DFG) - Project ID 423116110. 
This work was supported by the Cluster of Excellence Precision Physics, Fundamental Interactions, and Structure of Matter (PRISMA+ EXC 2118/1) funded
by the DFG within the German Excellence Strategy (Project ID 39083149); from the German-Israeli Foundation for scientific research and development (Grant \#1490); and by the Australian Research Council grants DP190100974 and  DP200100150, and the JGU Gutenberg Fellowship.

\bibliographystyle{apsrev4-1}
\bibliography{bibliography.bib}

\clearpage

\setcounter{table}{0}
\renewcommand{\thetable}{S\arabic{table}}%
\setcounter{figure}{0}
\renewcommand{\thefigure}{S\arabic{figure}}%
\setcounter{equation}{0}
\renewcommand{\theequation}{S\arabic{equation}}%

\section*{Supplemental material}
\section{Experiment}

\subsection{\label{sec:daq_anal}Data acquisition and analysis}

When the phase-modulation frequency $\nu_{\rm{m}}$ is tuned to nominally match $\delta\nu^{AA^{\prime}}/2$ between a pair of isotopes~($>$500~MHz), sweeping the laser frequency $\nu_{\rm L}$ in a small ($\sim$5\,MHz) range results in two independent time-traces, one for each isotope's spectrum, that are produced, respectively, by one of the frequencies present the laser light. 

The observed spectral profile of an isotope has a Full Width at Half Maximum (FWHM) of $\approx$\,800 kHz in atomic frequency (ie. twice the  width determined in laser frequency). This is  primarily determined by the transit time of atoms through the intracavity field of typical radius $w \approx$ 200\,$\rm{\mu}$m ($1/e^2$ intensity radius). The resulting transit-time broadening is $\Delta \nu_{tt}=(2/\pi) \sqrt{ln2}\,(\upsilon/w)\approx$\,770\,kHz for a thermal atomic velocity $\upsilon=\sqrt{2k_{\rm B}T/m}\approx290$\,m/s, at an Yb oven temperature $T=600\,\,^o$C, where $k_{\rm B}$ is the Boltzman constant and $m$ is the atomic mass. (Note that for a two-photon transition as studied here, this broadening is $\sqrt{2}$ times greater than that in a one-photon transition \cite{LehmannJCP2021}.) Aside from the finite lifetime of the excited state ($\tau=6.7\,\mu$s \cite{BowersPRA1996}), additional broadening may, in principle, arise due the finite width of the temporal profile of the intracavity field intensity ($\approx$50 $\mu s$) in the scanning Fabry-Perot cavity. As this width is substantially larger than the $\approx 1.3\, \mu s$ transit-time of atoms through the optical field, there is little additional broadening to be expected. That transit-time broadening is dominant in the observed spectra is supported by analysis of width ratios for different isotopes. From a set of analysed data on isotopes with A=172 and 174 we obtain a ratio $\rm {FWHM_{172}/FWHM_{174}=1.0062(5)}$, which is in agreement with the expected value based on transit-time:  $\Delta\nu_{tt}^{172}/\Delta\nu_{tt}^{174}=\upsilon_{172}/\upsilon_{174}=\sqrt{m_{174}/m_{172}}=1.0058$.

\begin{figure}[b]
    \centering
    \includegraphics[width=\columnwidth]{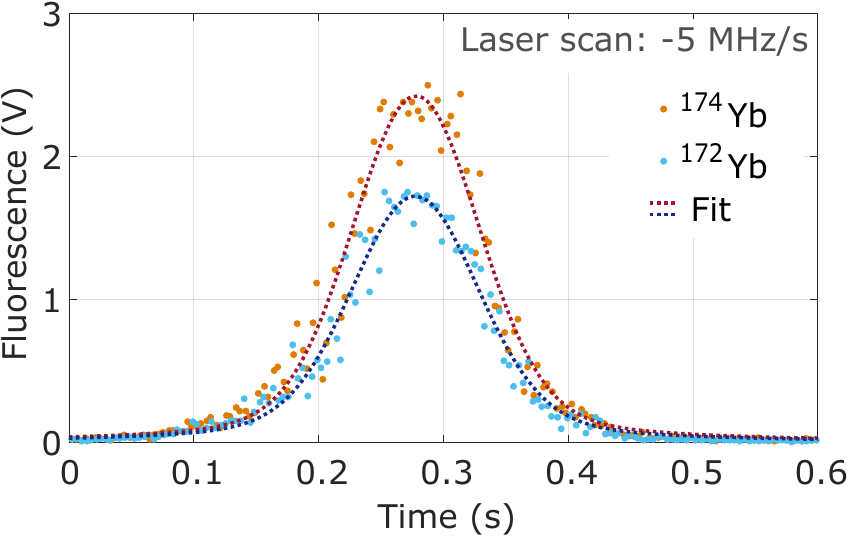}
    \caption{Example spectra on the \mbox{$m=0\rightarrow m'=0$} component of the $^1S_0 \rightarrow \, ^1D_2$ transition, and pseudo-Voigt profiles fitted to data.}
    \label{fig:SMPeak}
\end{figure}

In the absence of systematic effects, the IS for an isotope pair can be obtained by finding the frequency $\nu^0_{\rm{m}}$ for which the traces like the ones in Fig.~\ref{fig:SMPeak} are lined up.

To do this, each peak is fitted with a pseudo-Voigt function which is a linear combination of a Gaussian, $\mathcal{G}$ and a Lorentzian $\mathcal{L}$,
\begin{equation}
\label{eq:Lineshape}
    f(\nu) = \mu \mathcal{L}(\nu - \nu_0, \Gamma_L) + (1-\mu) \mathcal{G}(\nu- \nu_0, \Gamma_G),
\end{equation}
characterized by a mixing parameter $\mu \in$ [0,1]. This parameter, as well as the widths, $\Gamma_L$, $\Gamma_G$ and center $\nu_0$, are determined by least squares-fitting for each peak. The delay between a pair of peaks from different isotopes is sensitive to laser-scan imperfections, for example scan nonlinearity. To minimize this effect, we normalized the delay to the average FWHM, $\Gamma$, of both peaks,
\begin{equation}
    \Delta X = \frac{ \nu_0 - \nu_0'}{ \frac{1}{2}(\Gamma + \Gamma')}.
\end{equation}
This normalized delay $\Delta X$ will be zero when both isotopes are excited simultaneously as the laser is scanned. Because we are observing a two-photon transition, and in the absence of systematic effects, this will happen when the sideband is separated to the carrier by $\delta\nu^{AA^{\prime}}/2$.

Part of the delay depends on the sideband frequency, $\Delta X = \Delta X_0 + \alpha \nu_{\rm{m}}$, where $\Delta X_0$ and $\alpha$ are parameters to be determined from the data. In this case, $\delta\nu^{AA^{\prime}}/2$ can be obtained by determining $\nu_{\rm{m}}$ when the delay $\Delta X = 0$:
\begin{equation}
\label{eq:DeltaNu}
     \nu_{\rm{m}}\bigg\rvert_{\Delta X = 0} \equiv \nu^0_{\rm{m}} = -\frac{\Delta X_0}{\alpha}.
\end{equation}

\begin{figure}
    \centering
    \includegraphics[width=\columnwidth]{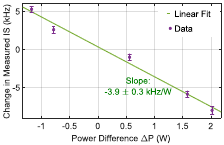}
    \caption{ \small{Systematic shift in the $\delta\nu^{AA^{\prime}}$ extraction due power imbalance between carrier (exciting $^{172}$Yb) and sideband field (exciting $^{174}$Yb). Because of the ac-Stark effect, individual lineshapes are shifted by  $\approx$20 kHz for the typical 5 W intracavity power in carrier or sideband fields.} }
    \label{fig:SMACStark}
\end{figure}

This is the essence of the method. However, there is another possible source for a delay: if the power of the carrier and relevant sideband are different by $\Delta P$, there will be a proportional shift/delay due to the ac-Stark effect. Indeed, while the ac-Stark shift is off-resonant, the peculiarity of our technique is that due the FP cavity scanning, the cavity is tuned to different components of the laser light at different times, so that the difference of power of the different components will result in a differential ac-Stark shift during the measurement of different peaks, leading to systematics. Figure \ref{fig:SMACStark} illustrates the impact of this effect on the $\delta\nu^{AA^{\prime}}$ extraction, in the presence of an exaggerated imbalance. In order to null the imbalance $\Delta P$ to within 0.1\%, we continuously measured the scanning FP transmission spectrum and adjusted the rf power driving the EOM. Additionally, during data taking we record the residual $\Delta P$ and include the resulting ac-Stark effect in the analysis via the free parameter $\beta$: 
\begin{equation}
\label{eq:PlaneFit}
    \Delta X = \Delta X_0 + \alpha \nu_{\rm{m}} + \beta \Delta P.
\end{equation}

\noindent We obtain a set of $\sim$200 $\Delta X$ values by recording and analysing spectra like those of Fig. \ref{fig:SMPeak} for different $\nu_{\rm{m}}$ (in the range $\pm$10 kHz around $\nu^0_{\rm{m}}$). 
We fit a plane as described by Eq.\,(\ref{eq:PlaneFit}) to the data and obtain a $\delta\nu^{AA^{\prime}}$ measurement through Eq.\,(\ref{eq:DeltaNu}). 

\begin{figure}
    \centering
    \includegraphics[width=\columnwidth]{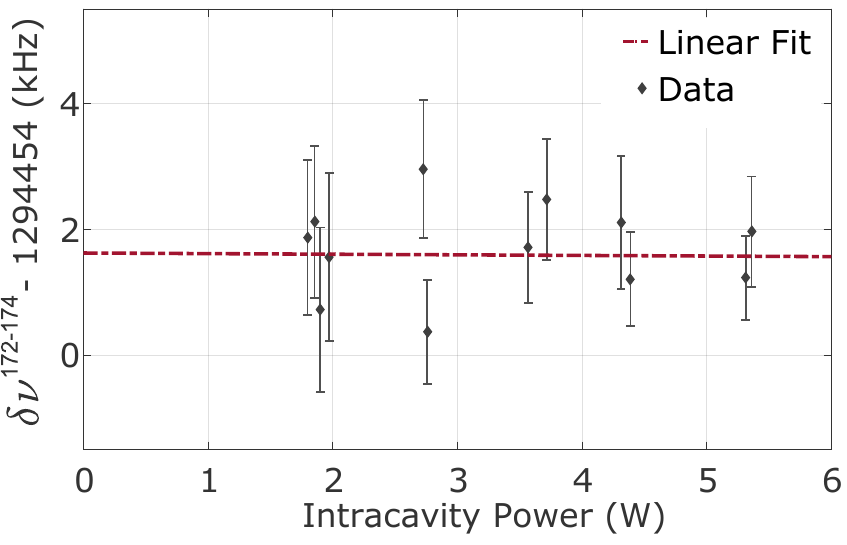}
    \caption{ \small{Extrapolation of measured $\delta\nu^{172-174}$ to zero intracavity power. The 12 data points represent $\delta\nu^{172-174}$ values measured between $^{172}$Yb (excited by carrier) and $^{174}$Yb (excited by sideband). 
    }}
    \label{fig:SMRound}
\end{figure}

As a final precaution against  residual  ac-Stark shifts, we take data at several intracavity powers and determine the value of $\delta\nu^{AA^{\prime}}$ at zero power through extrapolation. A total of 12 acquisition runs are made for  both laser scan directions and six intracavity power values,  yielding a total of 12 $\delta\nu^{AA^{\prime}}$ measurements. The extrapolation of $\delta\nu^{AA^{\prime}}$ to zero power (Fig.~\ref{fig:SMRound}) is expected to be free of residual ac-Stark-effect related systematics. In fact, our data from all  isotope pairs measured (a total of 113 data sets like that of  Fig.~\ref{fig:SMRound}) yield a mean slope of the straight line fitted to data which is consistent with zero. The determined value of  $-14(31)$~Hz/W implies no dependence of $\delta\nu^{AA^{\prime}}$ on intracavity power to within the measurement sensitivity.

A single $\delta\nu^{AA^{\prime}}$ determination from zero-power extrapolation has statistical accuracy on the order of 1 kHz. Many such $\delta\nu^{AA^{\prime}}$ extractions (16 to 25 extractions, depending on isotope pair) are made to reach the attained accuracy reported for each isotope pair on Table I of the main manuscript. We perform repeated data runs like that described above, to determine  $\delta\nu^{AA^{\prime}}$ values under different experimental conditions, as discussed in Section~\ref{sec:ComplemChecks}.

\subsection{\label{sec:SystEffects}Systematic effects}

\subsubsection{\label{sec:ParasBackgrounds} Parasitic signal backgrounds}
 
\begin{figure}
    \centering
    \includegraphics[width=\columnwidth]{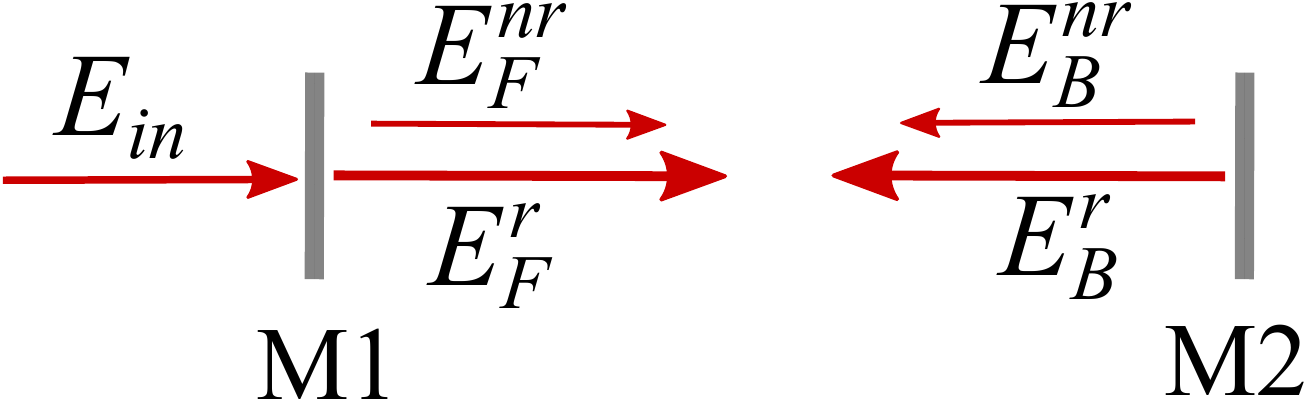}
    \caption{ \small{Intracavity fields  relevant for creation of parasitic signal backgrounds.  A field resonating between the FP mirrors (M1, M2) and propagating in the forward (F) or backward (B) direction, in combination with a non-resonant (nr) field propagating in the backward or forward direction, respectively, produces parasitic Doppler-free excitations. }}
    \label{fig:cavity-sys}
\end{figure}

\begin{figure}
    \centering
    \includegraphics[width=\columnwidth]{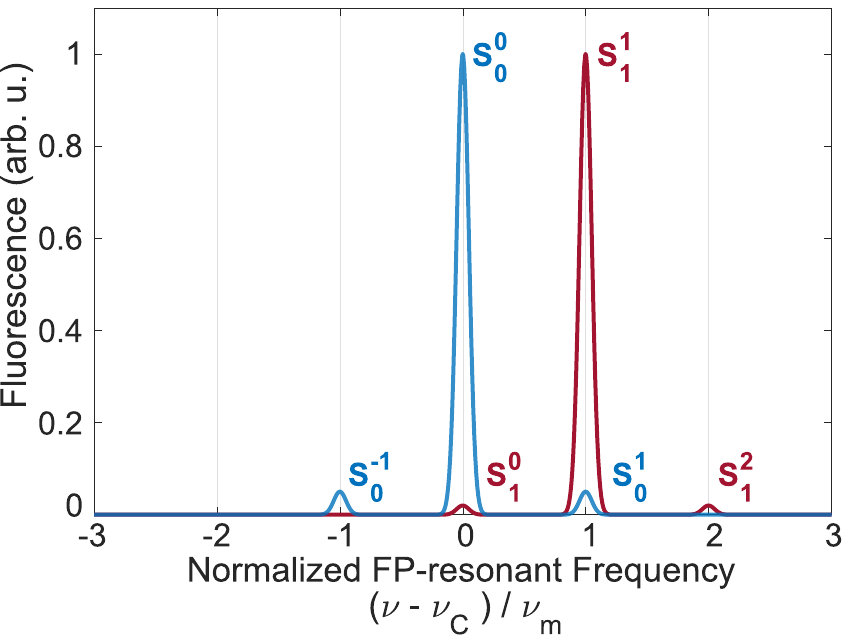}
    \caption{ \small{Contributions to the fluorescence signal in the various detection channels. Here we assume equal abundance for the two isotopes, primarily excited by carrier (blue, $\rm{n}=0$ channel) and first-order sideband (red, $\rm{n}=+1$ channel). Only the TEM$_{00}$ FP mode is considered.}}
    \label{fig:SMSpuriousIntro}
\end{figure}

The phase-modulated optical field incident on the FP input mirror is given by: 
\begin{equation}
\label{eq:SMEfield}
    E_{\rm{in}} = E_{0}\sum_{n=-\infty}^{\infty} J_{\rm{n}}(\delta)e^{i2\pi(\nu_{\rm{C}}-n\cdot \nu_{\rm{m}})t},
\end{equation}
 \noindent where $E_{\rm{0}}$ is the amplitude of the unmodulated field, $J_{\rm{n}}$ are Bessel functions of the first kind of order $n$, and $\delta$ is the modulation index. The index is set to $\delta\approx 1.44$, yielding equal powers ($\approx 30.0\%$ of total power) for the carrier and $1^{\rm{st}}$-order sideband fields, of frequency $\nu_{\rm{C}}$, $\nu_{\rm{C}} \pm \nu_{\rm{m}}$, respectively.  The remaining power is primarily in the sidebands with $n=\pm 2$ ($\approx 4.7\%$ each). Although the FP is resonant with a single frequency component at a given time, $\approx 3.5 \%$ of the power of the other components is still transmitted through the input mirror of the FP. The combination of the FP-resonant field with a counter-propagating, off-resonant frequency component (Fig.~\ref{fig:cavity-sys}) results in Doppler-free excitation of an isotope that is not excited by the FP-resonant field alone, creating a parasitic background (illustrated in Fig.~\ref{fig:SMSpuriousIntro}) that can lead to an erroneous measurement of $\Delta X$.

Considering this effect, the fluorescence signal $S^{\rm{n}}$ of a given detection channel $n$, i.e.  detection when the carrier frequency $\nu_{\rm{C}}$ ($n=0$) or one of the $1^{\rm{st}}$-order sidebands $\nu_{\rm{C}} \pm \nu_{\rm{m}}$ ($n=\pm1$) is resonant in the FP,  will have two contributions: $S^{\rm{n}} = S^{\rm{n}}_{\rm{n}} + S^{\rm{n}}_{\rm{i\neq n}} $ (Fig. \ref{fig:SMSpuriousIntro}). Here, $S^{\rm{n}}_{\rm{n}}$ is the primary contribution from the frequency component resonant with the FP. The contributions  $S^{\rm{n}}_{\rm{i\neq n}}$ arise from an isotope which is ideally only excited by some other frequency component with sideband index $i\neq n$. These additional contributions can affect the fitted centers and widths of the peaks, leading to an erroneous measurement of $\Delta X$. 
The $S^0_{+1}$ or $S^0_{-1}$ come from excitation caused by a combination of both $1^{\rm{st}}$-order sidebands, one which is FP-resonant and one which is not. Similarly, $S^{+ 1}_{0}$ or $S^{-1}_{0}$ come from the excitation caused by a non-resonant $2^{\rm{nd}}$-order sideband and the FP-resonant carrier. However, because the $2^{\rm{nd}}$-order sidebands have less power than that of the $1^{\rm{st}}$-order sidebands, the $S^{\pm 1}_{0}$ contribution is expected to be 
smaller than $S^0_{\pm 1}$. 

There are several factors that affect the characteristics of the parasitic backgrounds. For instance, if there is misalignment between the input field and the FP resonant mode axis, the non-resonant fields will propagate at an angle with respect to the resonant field, so the backgrounds will have a Doppler shift proportional to the misalignment angle. Additionally, this misalignment can lead to a smaller overlap region between the non-resonant and FP-resonant fields, resulting in a smaller background amplitude.

\begin{figure}
    \centering
    \includegraphics[width=\columnwidth]{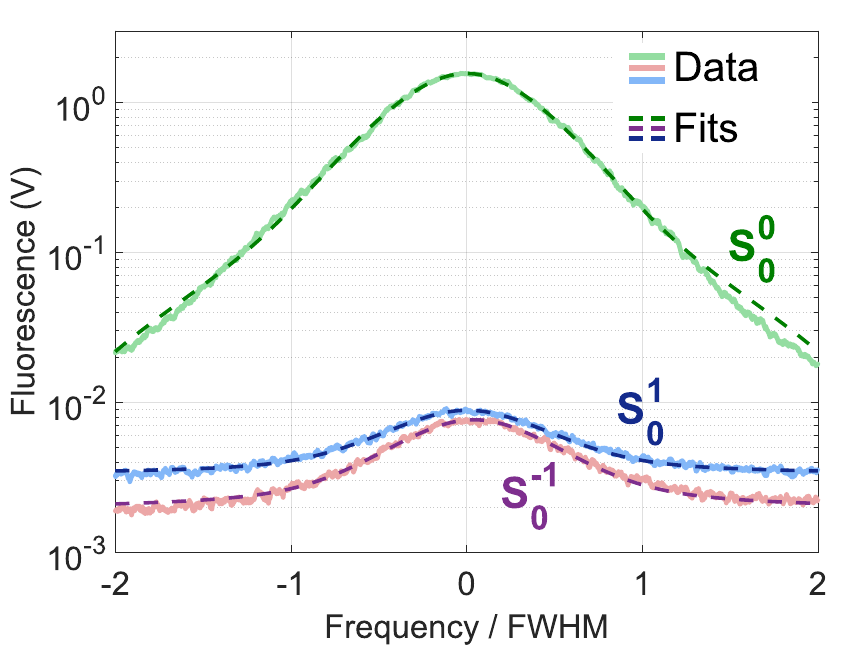}
    \caption{ \small{Measurement of the $S_{0}^{+ 1}$ and $S_{0}^{- 1}$ parasitic backgrounds, average of 60 scans. An isotope is excited with $2\nu_{\rm{C}}$, but $\nu_{\rm{m}}$ is detuned by $+3$\,MHz from $\delta\nu^{AA^{\prime}}$/2, so that the components $2(\nu_{\rm{C}} \pm \nu_{\rm{m}})$ do not excite any isotope. This allows for the characterization of $S_{0}^{\rm{+1}}$ and $S_{0}^{\rm{-1}}$, as the otherwise dominant $S_{+1}^{\rm{+1}}$ and $S_{-1}^{\rm{-1}}$ contributions, respectively, are now suppressed.}}
    \label{fig:SMParasitic}
\end{figure}

\begin{figure}
    \centering
    \includegraphics[width = \columnwidth]{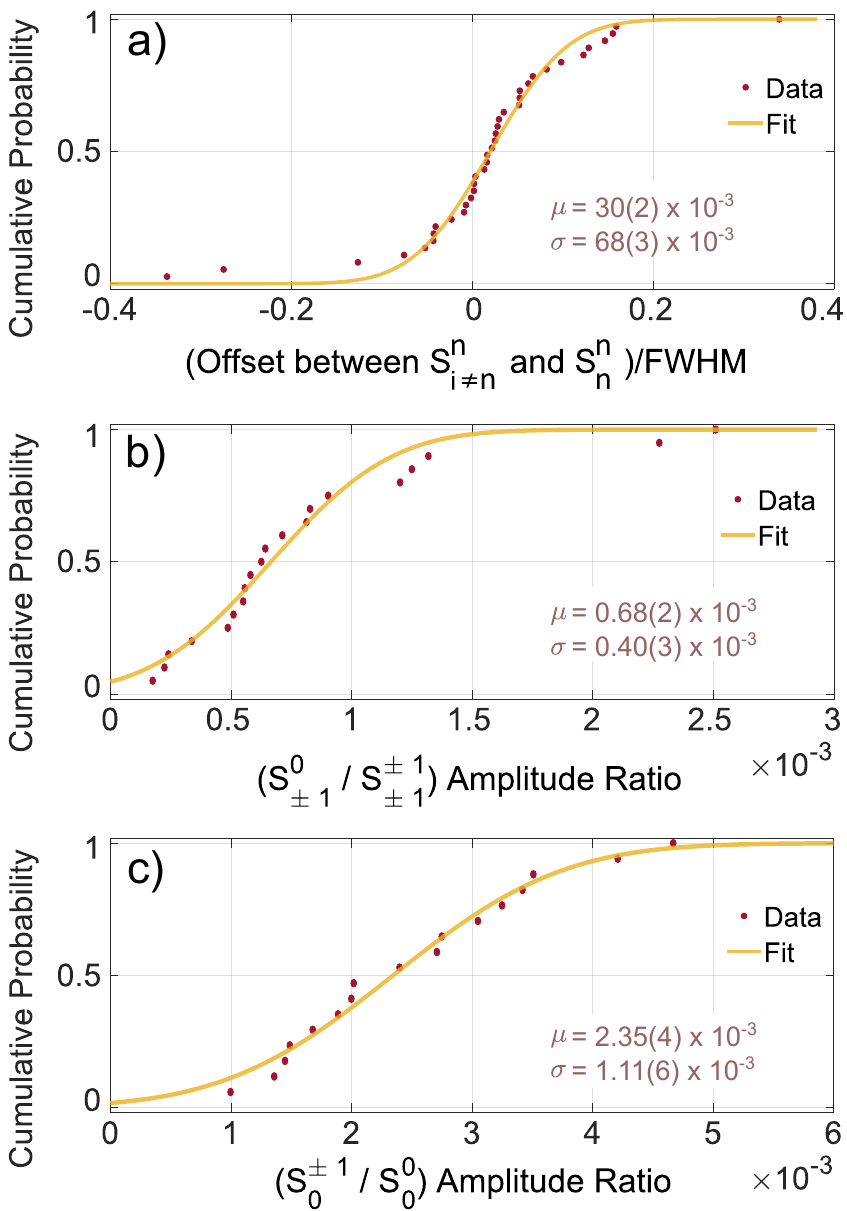}
    \caption{ \small{ Empirical cumulative distribution functions (CDF) for the offsets between $S^{\rm{n} }_{\rm{n \neq i}}$ and $S^{\rm{n = i}}_{\rm{n}}$ (a) and the $S^{0}_{\pm 1} /S^{\pm 1}_{\pm 1}$, and the $S^{\pm 1}_{0} /S^{0}_{0}$ amplitude ratios (b,c). A Gaussian CDF was used for fitting in all three cases. The mean offset in (a) corresponds to a misalignment of the incoming field with respect to the resonant mode of the FP of $\approx$50~$\mu$rad.} }
    \label{fig:SMSpuriousDistros}
\end{figure}

To characterize the parasitic backgrounds, we detuned $\nu_{\rm{m}}$ by $\pm 2-4$~MHz from $\delta\nu^{AA^{\prime}}/2$ so that only one isotope was excited during a laser scan. This allows us to study the backgrounds due to the excited isotope for a given cavity configuration. Then, the offsets and amplitude ratios between $S^{\rm{n}}_{\rm{i \neq n}}$ and $S^{\rm{n}}_{\rm{n}}$ can be obtained by fitting to the spectra, as shown in Fig.~\ref{fig:SMParasitic}. The distributions of the measured offsets between $S^{\rm{n}}_{\rm{i \neq n}}$ and $S^{\rm{n}}_{\rm{n}}$, and the $S^{\rm{n}}_{\rm{i \neq n}} /S^{\rm{n}}_{\rm{n}}$ amplitude ratios are shown in Fig.~\ref{fig:SMSpuriousDistros}.

This systematic was first observed after the two largest -abundance isotope pairs  $^{176}$Yb/$^{174}$Yb and $^{174}$Yb/$^{172}$Yb were measured, so the respective backgrounds were not measured and accounted for during the actual measurement runs for these pairs. (Instead, a posteriori studies of the systematic, as described below, were done to correct data for these pairs). This systematic is easier to observe for isotope pairs that have substantially different abundances because the influence of the parasitic backgrounds in the measurement of $\delta\nu^{AA^{\prime}}$ is larger, as the amplitude of the abundant isotope's background can be comparable to the signal of the less abundant isotope. For the $^{172}$Yb/$^{170}$Yb and $^{170}$Yb/$^{168}$Yb $\delta\nu^{AA^{\prime}}$ measurements, we characterized the parasitic backgrounds in the detection of the least abundant isotope of the pair, due to the larger-abundance isotope.  These measurements were done at the start and end of a daily experimental run.  We subtracted the background in the spectrum of the less abundant isotope by scaling and shifting the fitted spectrum of the more abundant isotope by the measured characteristic amplitude ratio and offset of the background. By comparing the results (after the extrapolation to zero power) with and without background subtraction, we obtained a value for the correction for a given isotope pair and experimental run (e.g. a run with a particular isotope excited by carrier and another by sideband). An uncertainty was assigned to each correction value and added in quadrature to the statistical uncertainty of a $\delta\nu^{AA^{\prime}}$ determination from extrapolation to zero power (see Fig. \ref{fig:SMRound}). This uncertainty in the correction was calculated considering the uncertainties in the measured amplitude ratios and offsets of the backgrounds.

As a given isotope pair was measured many times over different days, a set of parasitic background corrections was produced for a given pair.  Each set yields a weighted mean for the correction that is shown in Table~\ref{tab:SMCorrectionTable} as $\Delta (\delta\nu^{AA^{\prime}})_{\rm{meas}}$ along with its standard error.

However, the aforementioned background treatment does not account for all the effect, as there are situations in which the influence of the less abundant isotope's background is comparable to that of the more abundant isotope; for example in the case where the isotopic abundances are similar and the less abundant isotope is excited by the carrier frequency, as $S_0^{\pm1}/S_0^{0}$ is $\approx$3.5 times larger than $S_{\pm1}^0 / S_{\pm1}^{\pm1}$, as shown in Fig.~\ref{fig:SMSpuriousDistros}. In order to account for these contributions that were not treated during the actual IS measurement runs, we took the following two steps. First, we carried out dedicated data runs under changing experimental conditions (i.e. with varying FP length and input-beam alignment), to thoroughly study the parasitic background amplitudes and offsets and expand the available set of data as to obtain reliable distributions of their occurrences. These distributions are shown in Fig.~\ref{fig:SMSpuriousDistros}. These auxilliary runs were made using the $^{174}$Yb/$^{176}$Yb pair. Second, we carried out posteriori Monte Carlo (MC) simulations of the effect, based on the measured distributions of Fig.~\ref{fig:SMSpuriousDistros}.
 This allowed us to obtain a correction on $\delta\nu^{AA^{\prime}}$ for the cases where the effect was partially subtracted ($^{172}$Yb/$^{170}$Yb and $^{170}$Yb/$^{168}$Yb), and the cases where none of the backgrounds were characterized ($^{176}$Yb/$^{174}$Yb and $^{174}$Yb/$^{172}$Yb).

The Monte Carlo procedure consisted of simulating lineshapes as described by Eq.~(\ref{eq:Lineshape}) with a relative shift of $\Delta X$ and an amplitude ratio given by the natural abundances of the isotopes to be considered. Parasitic backgrounds were added to these lineshapes, with their offsets and amplitude ratios being sampled from the measured distributions for these parameters, shown in Fig.~\ref{fig:SMSpuriousDistros}. Then, $\Delta X'$ was obtained with the presence of parasitic backgrounds by fitting to the spectra. After repeating this for different values of $\Delta X$, we extracted the intercept $\Delta X \rvert_{\Delta X' = 0}$, which corresponds to the correction in the $\delta\nu^{AA^{\prime}}$ measurement times $\alpha/2$, where $\alpha$ is the fit parameter of Eq.~(\ref{eq:PlaneFit}). With this method we calculated the corrections to the $\delta\nu^{AA^{\prime}}$ measurements when both ($S_0^{\pm1}$ and $S_{\pm1}^0$) parasitic backgrounds are considered ($\Delta(\delta\nu^{AA^{\prime}})_{\rm{MC}}$). Additionally, we calculated the contributions when only one parasitic background was added: when only $S_0^{\pm1}$ contributions are considered, the correction to $\nu^{AA^{\prime}}$ is $\Delta(\delta\nu^{AA^{\prime}}_{0\rightarrow \pm1})_{\rm{MC}}$, and when only considering $S_{\pm1}^0$ contributions, there is a correction of $\Delta(\delta\nu^{AA^{\prime}}_{\pm1 \rightarrow 0})_{\rm{MC}}$. The values for these corrections are presented in Table~\ref{tab:SMCorrectionTable}. The corrections were applied to the data in different ways: for isotope pairs $^{176}$Yb/$^{174}$Yb and $^{174}$Yb/$^{172}$Yb, the full correction $\Delta(\delta\nu^{AA^{\prime}})_{\rm{MC}}$ was added to every $\nu^{AA^{\prime}}$ measurement (after extrapolation to zero power). For isotope pairs $^{172}$Yb/$^{170}$Yb and $^{170}$Yb/$^{168}$Yb however, large part of the effect was already considered by the subtraction procedure, so only the missing part of the correction (the contribution of the less-abundant isotope in the more-abundant isotope's background), was applied in these cases. This was done in the same way as before: by adding corrections to every $\nu^{AA^{\prime}}$ measurement (after extrapolation to zero power). The corrections that were added this way are highlighted in Table~\ref{tab:SMCorrectionTable}. We assumed a conservative 100\% uncertainty in the correction, which was added in quadrature to the uncertainty of every individual $\delta\nu^{AA^{\prime}}$ measurement after the extrapolation to zero power. As these added uncertainties are smaller than the typical $\sim$1 kHz statistical uncertainty of an individual $\delta\nu^{AA^{\prime}}$ determination through zero-power extrapolation (Fig. \ref{fig:SMRound}), there is negligible increase of overall uncertainty in the final results.
 
\begin{table*}[htb]
    \centering
    \caption{ Influence of parasitic backgrounds on $\delta\nu^{AA^{\prime}}$. The isotope that was excited by the carrier frequency is indicated by a C in the subscript. The data in bold font were used to apply corrections to the measured IS. The numbers in  in parenthesis indicate the standard error of the mean. Whenever an MC-simulated correction was applied to a measured $\delta\nu^{AA^{\prime}}$ value, a conservative 100\% error was assumed in the correction. $\Delta (\delta\nu^{AA^{\prime}})_{\rm{meas}}$ is the measured difference in $\delta\nu^{AA^{\prime}}$ after subtracting the parasitic background due to the more abundant isotope, as described in the text.}
        \begin{ruledtabular}
            \begin{tabular}{ c | c c c | c}

                \textrm{Isotope pair (A$_{\rm{C}}$-A$^{\prime}$)}&
                \textrm{$\Delta(\delta\nu^{AA^{\prime}})_{\rm{MC}}$ (Hz)}&
                \textrm{$\Delta(\delta\nu^{AA^{\prime}}_{0\rightarrow \pm1})_{\rm{MC}}$ (Hz)} &
                \textrm{$\Delta(\delta\nu^{AA^{\prime}}_{\pm1 \rightarrow 0})_{\rm{MC}}$ (Hz)} &
                \textrm{$\Delta (\delta\nu^{AA^{\prime}})_{\rm{meas}}$ (Hz)}\\
                \colrule
                176$_{\rm{C}}$-174& \textbf{27(4)}& -22(1)& 45(4)&-\\
                174$_{\rm{C}}$-176& \textbf{-144(11)}& -147(7)& 8(1) &-\\
                174$_{\rm{C}}$-172& \textbf{-77(3)}& -88(6)& 13(1)&-\\
                172$_{\rm{C}}$-174& \textbf{-17(3)}& -40(3)& 27(1)&-\\
                172$_{\rm{C}}$-170& -340(24)& -329(16)& \textbf{2(1)}&-1052(668)\\
                170$_{\rm{C}}$-172& 106(11)& \textbf{-6(1)}& 104(10)&38(14)\\
                168$_{\rm{C}}$-170& 365(8)& \textbf{-2(1)}& 374(12)& 55(72)\\

            \end{tabular}
        \end{ruledtabular}
    \label{tab:SMCorrectionTable}
\end{table*}

While  parasitic backgrounds have more prominent influence when considering measurements using carrier and $1^{\rm{st}}$ order sideband fields, such backgrounds  were also observed in the experiments were both isotopes were excited with the two 1$^{\rm{st}}$-order sidebands. These contributions arise due to the combined excitation caused by the FP-resonant $1^{\rm{st}}$-order sideband and a non-resonant $3^{\rm{rd}}$-order sideband, which carries a small, but nonnegligible power 0.3\% of the total. The measured amplitude ratio for these parasitic contributions, $S_{\mp1}^{\pm1} / S_{\pm1}^{\pm1}$ had a mean value of $0.7(2)\times 10^{-5}$, two orders of magnitude smaller than the amplitude ratios observed for the experiments with an isotope being excited with the carrier. For the $^{168}$Yb/$^{172}$Yb isotope pair, the mean correction on $\delta\nu^{AA^{\prime}}$ (considering only the background due to $^{172}$Yb) was $11(13)$~Hz, which should account for the effect given the large disparity in abundances. 

The backgrounds for the $^{172}$Yb/$^{176}$Yb pair were not measured, but their effect is practically negligible as the abundance ratio for this pair is closer to unity. Indeed, a MC simulation of the expected shift in this case, yields a value of $3(1)$~Hz, which is too small to be of concern, given the $339$~Hz accuracy achieved in this complementary measurement (i.e. a measurement not used in King-plot analysis).  

 \subsubsection{\label{sec:NLlaserscan} Nonlinearity of laser scan}

\begin{figure}
\includegraphics[width=\columnwidth]{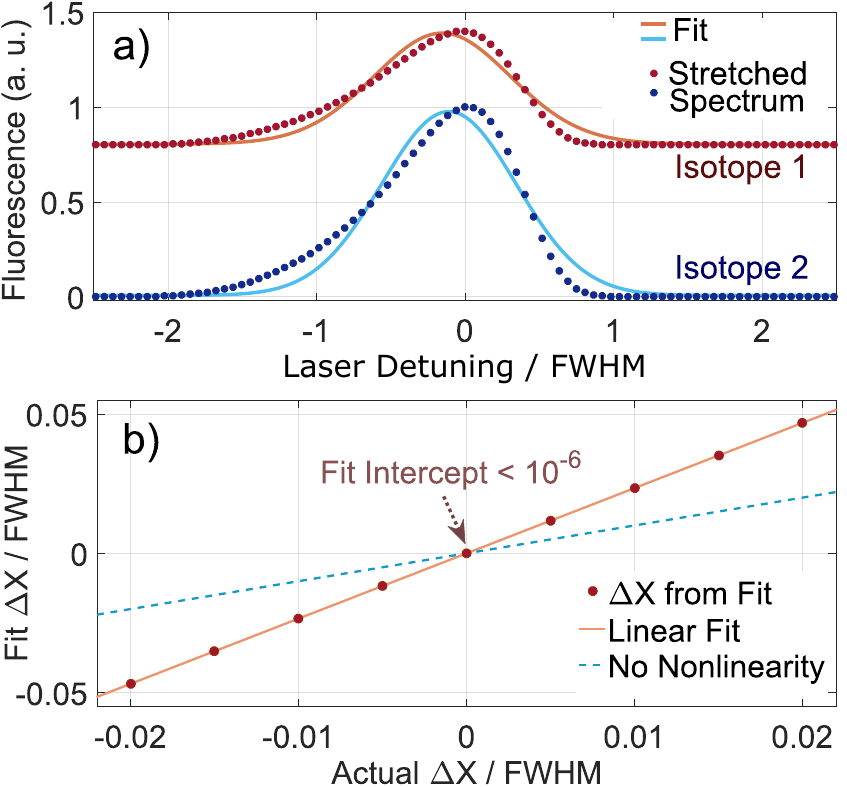}
\caption{\small{ Effect of laser scan nonlinearity on $\delta\nu^{AA^{\prime}}$ determination. a) Distorted lineshapes (vertically offset for clarity) and pseudo-Voigt fits to these. b) Extracted delay between the two isotope spectra, plotted against the delay in the absence of nonlinearity.  }}
\label{fig:laserscanNL}
\end{figure}

 The simultaneous excitation of two isotopes serves to largely eliminate systematics due to laser-scan nonlinearity. Here we simulate spectra to study the residual impact of this nonlinearity on the $\delta\nu^{AA^{\prime}}$ extraction and show that it is negligible.
 Figure \ref{fig:laserscanNL}a shows simulated lineshapes in the presence of an exaggerated nonlinearity, resulting in pronounced lineshape asymmetry. To produce such a nonlinearity, a stretching transformation is applied to the frequency axis. The  solid lines represent pseudo-Voigt fits to the spectra, that are used to extract the relative delay $\Delta X$ between the distorted peaks. This delay is plotted in Fig. \ref{fig:laserscanNL}b against the delay in the absence of nonlinearity, i.e. the delay imposed due to the stepping of the $\nu_{\rm{m}}$ frequency. The stepping is done in a range $\pm$0.02 FWHM, similarly to the actual experiment.  As seen from a linear fit to the data of Fig. \ref{fig:laserscanNL}b, the y-intercept (corresponding to the zero-delay in the presence of nonlinearity) is practically null (it is less than $10^{-6} $ FWHM or $ \approx$1 Hz in actual frequency, and determined by numerical errors). The blue dotted line shows the dependence in the absence of nonlinearity. While significant systematic shifts arise away from the zero-delay point in the presence of nonlinearity, when data are taken about the nominal zero-delay (as is done in the actual acquisition) and extraction of $\delta\nu^{AA^{\prime}}$ is done by obtaining the intercept, we find a negligible residual systematic due to imperfections in the laser scan.
  
\subsubsection{\label{sec:Resid1stDoppler}Residual 1$^{\rm{st}}$-order Doppler effect}

While the observed experimental signal is primarily due to Doppler-free excitations from the two counter-propagating intracavity fields, a Doppler-broadened profile due to two-photon absorption from the same intracavity field  is also detectable. Its amplitude (width) is approximately $\times 70$ times smaller (larger) relative to that of the Doppler-free resonance. If the FP and atomic beam axes are not perpendicular to each other, the two background profiles  from the counter-propagating fields will be offset in frequency by $4(\upsilon_{\rm{}}/\lambda)\theta$, where $\upsilon \approx$\,290 m/s  is the atomic velocity and the angle $\theta$ represents the deviation from perpendicularity between the FP and atomic beam axes. Given the finite ($>99.5\%$) reflectivity of the FP output mirror  (M2 in Fig. \ref{fig:cavity-sys}), the two counter-propagating fields and thus, the respective background signals have slightly different amplitudes.  Thus, the total Doppler-broadened background will be offset in frequency with respect to the Doppler-free peak. This will result in a systematic shift in the extraction of the position of the fitted lineshape. While most of the shift will be common in two peaks from different isotopes, a small, residual shift is expected in $\delta\nu^{AA^{\prime}}$ because of the slight ($\approx$ 0.5\%) variation of the Doppler-shift between isotopes of adjacent mass number. 

Simulations of fluorescence signal including the Doppler-broadened background, carried out for a conservative value of $\theta=50$ mrad indicate small shift in the extraction of the single-isotope peak center ($<1$ Hz). Moreover, the impact on the $\delta\nu^{AA^{\prime}}$ extraction is two orders of magnitude smaller, and thus negligible given the uncertainties in the present $\delta\nu^{AA^{\prime}}$ determinations.

  \subsubsection{\label{sec:2ndorderDoppler}2$^{\rm{nd}}$-order Doppler effect}
  
  The 2$^{\rm{nd}}$-order Doppler effect causes a relative shift to the laser frequency in the atom frame of $d\nu_{\rm{L}}=\nu_{\rm{L}}(-\upsilon^2/2c^2)\approx -194$ Hz, where $\upsilon\approx 290$ m/s is the atomic velocity, and $c$ is the speed of light. The shift in the two-photon transition frequency is $\approx-388$ Hz. However, only its variation from isotope to isotope is relevant for the $\delta\nu^{AA^{\prime}}$ extraction. This variation is related to the $\approx$0.58\% velocity change between two isotopes of adjacent mass number ($\upsilon= \sqrt{2 k_B T/m}$).
  The residual shift is $\approx 4 $ Hz, for two isotopes with $\Delta A=2$. We apply corrections to the measured IS values to account for this effect.
%
  
  %
  
  \subsubsection{\label{sec:ResZeemanShifts} Effect of Zeeman shifts}
  
  In the absence of hyperfine interaction in the nuclear-spin-zero isotopes studied here, there is no 2$^{nd}$ order Zeeman effect to contribute to the measured $m=0\rightarrow m’=0$ component of the $^1S_0\rightarrow ^1D_2$ transition. 

However magnetic components of the transition, that appear parasitically in the spectrum due to a misalignment of the optical-field polarization axis with respect to the B=4\,G applied magnetic field, and due to a slight ellipticity of the optical field,  while well separated in the spectrum from the 0-0 component, may “pull” the observed   0-0 transition lineshape. $\Delta m=\pm 1$ components are observed in the spectrum with amplitudes of a few percent of the 0-0 peak, that are generally different for the two parasitic peaks. We have simulated the lineshape-pulling effect on the 0-0 component and see that it amounts to  20\,Hz, assuming a reasonable parasitic peak amplitude of 4\% (relative the 0-0 peak height)  for a $\Delta m=-1$ lineshape and 2\%  for the $\Delta m=+1$ lineshape. Since this pulling, however, is common to the 0-0 lineshapes of a pair of isotopes, it does not contribute to the measured isotope shift. 

\subsubsection{\label{sec:nmFrequencyAccuracy} $\nu_{\rm{m}}$ frequency accuracy}
     The frequency $\nu_{\rm{m}}$ is set with a signal generator used to drive the EOM, providing phase modulation of the laser light. The generator (SRS model SG386) is referenced to a GPS-disciplined time source (Meinberg model LANTIME M600/GPS), and the generator frequency is monitored with a frequency counter (SRS model SR620), that is also referenced to the same GPS-disciplined time source. The counter reading of $\nu_{\rm{m}}$ was checked and found to have actual offset from set value which was as large as 4 Hz. We correct the IS measurements of Table~I in the main paper accordingly   and assign a conservative 100\% uncertainty (8 Hz in the IS data) to these corrections. 
     

 \subsection{\label{sec:ComplemChecks}Additional experimental checks}

Data taking to extract $\delta\nu^{AA^{\prime}}$ was carried out under a variety of different conditions. This is done to ensure consistency of measurements and the check for unaccounted for systematic effects.
 
The most frequent change from run to run is alternating the use of the 1$^{st}$-order sideband. We take half the data with either sideband resonant with an isotope. This is primarily done to  check against potential residual ac-Stark effect due to parasitic mode interferences. Such interferences may arise, for example, if a higher-order TEM mode has resonance partially overlapping in the spectrum with the fundamental TEM$_{00}$ mode, that is ideally the only mode resonant in the FP. Switching between the two sidebands also allows to check consistency of data against systematic shifts due to parasitic background signals from off-resonant fields present in the FP, as discussed in the previous section.

Another potentially impactful change during runs is in the FP cavity length. We take data for a given isotope pair for at least two different cavity lengths. These lengths, depending on the isotope pair being measured, vary in the range 2-5 cm.  The primary motivation to take the FP apart and reassemble it at a different length is to ensure that parasitic cavity interferences which depend on the given FP mode structure (determined by the cavity length and the mode matching to various cavity modes) are sufficiently suppressed. Taking data with different FP length serves also to test against unexpected residual $1^{st}$-order Doppler effects in $\delta\nu^{AA^{\prime}}$, which, if present, may appear in the data when the FP axis is shifted with respect to the atomic beam axis. To within the statistical accuracy attained in each FP configuration, the measured $\delta\nu^{AA^{\prime}}$ are consistent with each other, as indicated by the results shown in Table \ref{tab:SMDiscrepancyTable}.

\begin{table}
    \centering
     \bigskip
\caption{Difference in the mean $\delta\nu^{AA^{\prime}}$ values between data subsets acquired for different experimental conditions (in standard deviations). In the C-S reversal test, the IS measurements with one isotope (A) excited by carrier and the other (\rm{A}$^\prime$) with a sideband, are compared with measurements with A  measured with a sideband and  A$^\prime$ with the carrier. N/A means that the test was not done.   
}


        \begin{ruledtabular}
            \begin{tabular}{ c c c c }

                \textrm{Isotope pair (A-A$^{\prime}$)}&
                \textrm{FP Length}&
                \textrm{Temperature}&
                \textrm{C-S Reversal}\\
                \colrule
                176-174& 0.30& 1.35& 0.62\\
                174-172& 0.16& 0.64& 1.46\\
                172-170& 1.87& 0.99& 0.66\\
                170-168& 0.34& N/A& N/A\\
                172-176& 0.56& 1.37& N/A\\
                168-172& 0.69& N/A& N/A\\
            \end{tabular}
        \end{ruledtabular}
    \label{tab:SMDiscrepancyTable}
\end{table}

In addition to varying the FP length 
we take data at different oven temperatures. This is a safeguard against unanticipated residual $1^{st}$-order Doppler effect in $\delta\nu^{AA^{\prime}}$, which could arise due to the slight atomic velocity variation ($\approx 0.5$\%) between adjacent-mass isotopes. We vary the temperature between T=500 and 600 $^o$C and obtain an equal amount of data at both temperatures (with the exception of data involving the $^{168}$Yb isotope, for which because of the its low abundance we only measured at 600 $^o$C). The measurements at different T are consistent within $\approx 1.4\sigma$ or less, as shown in Table \ref{tab:SMDiscrepancyTable}. 

\begin{figure*}
\includegraphics[width=2\columnwidth]{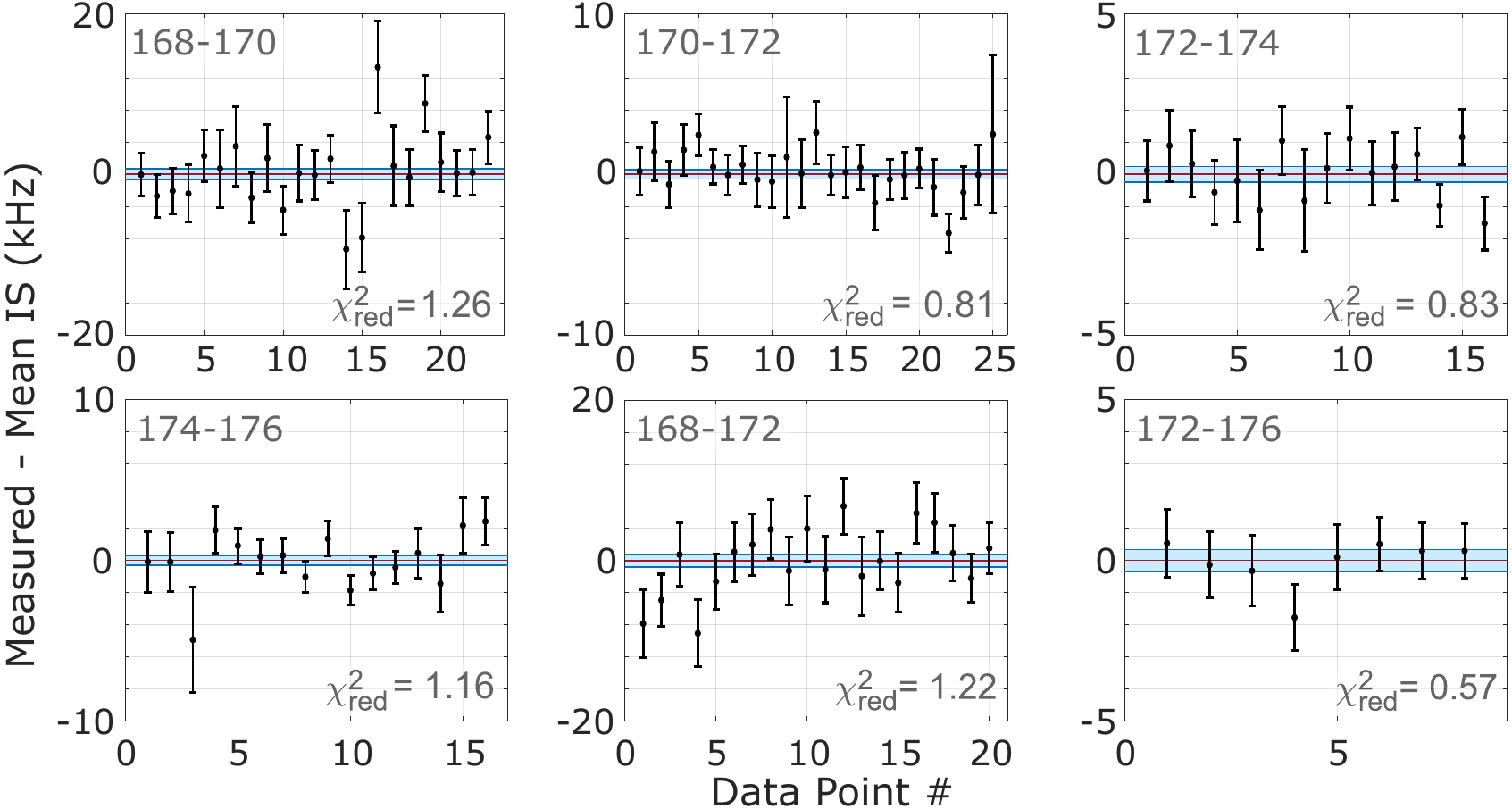}
\caption{\small{Summary of experimental data from all the runs.  The blue-shaded region corresponds to $\pm 1 \sigma$. The mean values of the respective determinations are listed in Table~I of the main manuscript}}
\label{fig:SMFinalSix}
\end{figure*}

The data from all runs carried out for each isotope pair are shown in Fig. \ref{fig:SMFinalSix}. Each data point is the result of extrapolating 12 $\delta\nu^{AA^{\prime}}$ measurements to zero power, as illustrated in Fig.~\ref{fig:SMRound}. Corrections to the data points for treatment of the parasitic background signals discussed in section B are included. The reduced $\chi^2$ values observed in the data sets indicate reasonable consistency of the individual $\delta\nu^{AA^{\prime}}$ determinations.

\section{Theory}

\subsection{\label{sec:CalculationIS}Atomic structure calculations}

Here we follow our previous work on isotope shift (IS) in Yb$^+$~\cite{AllehabiPRA2021}. We directly use the results of Ref.~\cite{AllehabiPRA2021} for Yb$^+$ and perform similar calculations for neutral Yb. 
It was demonstrated in Ref.\,\cite{AllehabiPRA2021} that the analysis of the King plot and its nonlinearities (NL) is sensitive to numerical noise. 
One way to suppress the noise is to perform the calculations for both transitions needed for the King plot in a similar way and to use the so-called random-phase approximation (RPA). The RPA method is linear with respect to the perturbation, which in our case is variously the change of nuclear potential due to the change of nuclear size $F$ or deformation $G^{(4)}$, or the potential due to a new boson $D$. By using the RPA method we separate the small correction to the energy from its large zero-order value, so that the uncertainties in the zero-order energies do not affect the calculations. The use of the RPA method also removes higher-order corrections to the energy. Therefore, the quadratic term containing the $G^{(2)}$ coefficient should be added separately.

In this work we study the isotope shifts in the \mbox{$^1\textrm{S}_0 \rightarrow\,^1\textrm{D}_2$} transition (`c') in neutral Yb and the $^2\textrm{S}_{1/2}\rightarrow \,^2\textrm{D}_{5/2}$ transition (`a')  in Yb$^+$~\cite{CountsPRL2020}.
Yb$^+$ has one valence electron while neutral Yb has two valence electrons. To make the calculations for both transitions similar, we reduce the calculations for Yb to the simpler task of calculating for an one-valence-electron system. To do this we start from the relativistic Hartree-Fock (RHF) procedure for Yb$^{2+}$. We then calculate the $6s$, $5d_{3/2}$ and $5d_{5/2}$ states of the valence electron in the field of the frozen core. These first two steps are exactly the same as in Yb$^+$ calculations. We then recalculate the valence states by adding to the potential the contribution of the $6s$ electron. Since all three states are recalculated, including the $6s$ state, the additional potential changes. Therefore, the process needs to be repeated until there is no change. Less than ten iterations are needed for full convergence. The CI calculations show that the $^1$D$_2$ state of Yb is composed of 20\% $6s5d_{3/2}$ configuration and 80\% $6s5d_{5/2}$ configuration.
Therefore, the $5d_{5/2}$ state, calculated in the field of frozen Yb$^{2+}$ core plus the field of the $6s$ electron can be approximately associated with the $^1$D$_2$ state of Yb. 

To improve the accuracy of the calculations we construct Brueckner orbitals (BO) for the valence states of Yb using the same correlation potential operator $\Sigma$ as for the Yb$^+$ ion ~\cite{AllehabiPRA2021}. We rescale the operator to fit the energies of the calculated $6s$, $5d_{3/2}$ and $5d_{5/2}$  states to the experimental energies of the $^1$S$_0$,  $^3$D$_2$, and $^1$D$_2$ states of neutral Yb.

To calculate the energy shift due to the change of nuclear size and shape we use the RPA method in exactly the same way as for Yb$^+$,
\begin{equation}
\label{e:FIS}
\delta\nu_i = \langle \psi_i^{\rm BO} | \delta V_N + \delta V_{\rm core}| \psi_i^{\rm BO} \rangle.
\end{equation}
Here, $\delta V_N$ is the difference between nuclear potentials for the two isotopes,  $\delta V_{\rm core}$ is the change of the self-consistent RHF potential induced by $\delta V_N$ and  the  changes to all core functions. These two terms are exactly the same as in Yb$^+$. The $\psi_i^{\rm BO}$ is the BO for the valence state of Yb$^+$ or Yb.

In the end our calculations for Yb$^+$ and Yb we have the same atomic core, the same RPA operator \mbox{$ \delta V_N + \delta V_{\rm core}$} and the same correlation potential $\Sigma$. The only difference is an additional contribution of the $6s$ electron to the HF potential for valence electrons.
This makes the calculations sufficiently similar to suppress numerical noise. The calculated slope of the King plot differs from the experimental value by only about 0.2\%, which is similar to what we had for two transitions in Yb$^+$ \cite{AllehabiPRA2021}.

As was mentioned above, the RPA equations are linear in the perturbation, therefore they do not include quadratic terms containing the $G^{(2)}$ coefficients. To calculate them, we use a second-order perturbation theory similar to what we have done for Yb$^+$ \cite{AllehabiPRA2021}.

In Table\,\ref{t:fit}, we present \emph{ab initio} calculations of the electronic structure factors from Eq.~(2) in main paper. The accuracy of the calculations for the parameters $F$ is  about 1\%, while it is about 10\% for $G^{(2)}$ and $G^{(4)}$. However, the accuracy for the ratios tends to be much higher. For example, the values of $G^{(2)}$ for the two transitions in Yb$^+$ are the same to five digits, due to dominating contribution from the $6s$ state and almost negligible contributions from the $5d_{3/2}$ and $5d_{5/2}$ states.

\begin{table}
	\caption{Calculated electronic isotope shift factors of Eq.\,(2) in main paper, for transitions of Yb$^+$ and Yb. See text for discussion of uncertainties.}
	\label{t:fit}
	\begin{ruledtabular}
		\begin{tabular}{lc ddd}
			\multicolumn{1}{c}{Index} &
			\multicolumn{1}{c}{Transition} &
			\multicolumn{1}{c}{$F$} &
			\multicolumn{1}{c}{$G^{(2)}$} &
			\multicolumn{1}{c}{$G^{(4)}$} \\
			\multicolumn{1}{c}{} &
			\multicolumn{1}{c}{} &
			\multicolumn{1}{c}{GHz/fm$^2$} &
			\multicolumn{1}{c}{GHz/fm$^4$} &
			\multicolumn{1}{c}{GHz/fm$^4$} \\
			\hline
			
			$a$ &  Yb$^+$ $6s$ -- $5d_{5/2}$  &-17.604 &  0.02853 & 0.01308 \\
			$b$ &  Yb$^+$ $6s$ -- $5d_{3/2}$  &-18.003 &  0.02853 & 0.01337 \\
			$c$ &  Yb $^1\textrm{S}_0$ -- $^1\textrm{D}_2$  &-14.437   &  0.02334 & 0.01042 \\
		\end{tabular}
	\end{ruledtabular}
\end{table}

\subsection{Effect of new boson}

The generalised King analysis used in the main text requires that the contribution of new physics be calculated with good accuracy. The term responsible for the new interaction [see Eq.~(2) in main paper] is $ ({\alpha_{\rm NP}}/{\alpha}) D_i \gamma^{AA^\prime}$, where $\alpha_{\rm NP}=q_nq_e/\hbar c$,  $q_n$ and $q_e$ are unknown coupling constants for neutrons and electrons, $\alpha = e^2/\hbar c$ is the fine structure constant, $D$ is the matrix element of Yukawa-type interaction, 
$$D=e^2 \left< \frac{\exp(-m_\phi cr/\hbar)}{r} \right>,$$
$m_{\rm \phi}$ is the boson mass, and $\gamma^{AA^\prime}$ is the difference in neutron number between two isotopes ($\gamma^{AA^\prime}=2$ in this work). 

The electronic structure factor $D$ depends on the mass of the new boson and must be calculated. 
We perform the calculations in the range $10^{-5}$ to $10^5$ keV using the RPA method and BO for both Yb and Yb$^+$. The Yukawa potential serves as a perturbation in the RPA calculations.

\begin{figure}[tb]
	\includegraphics[width=8.7cm]{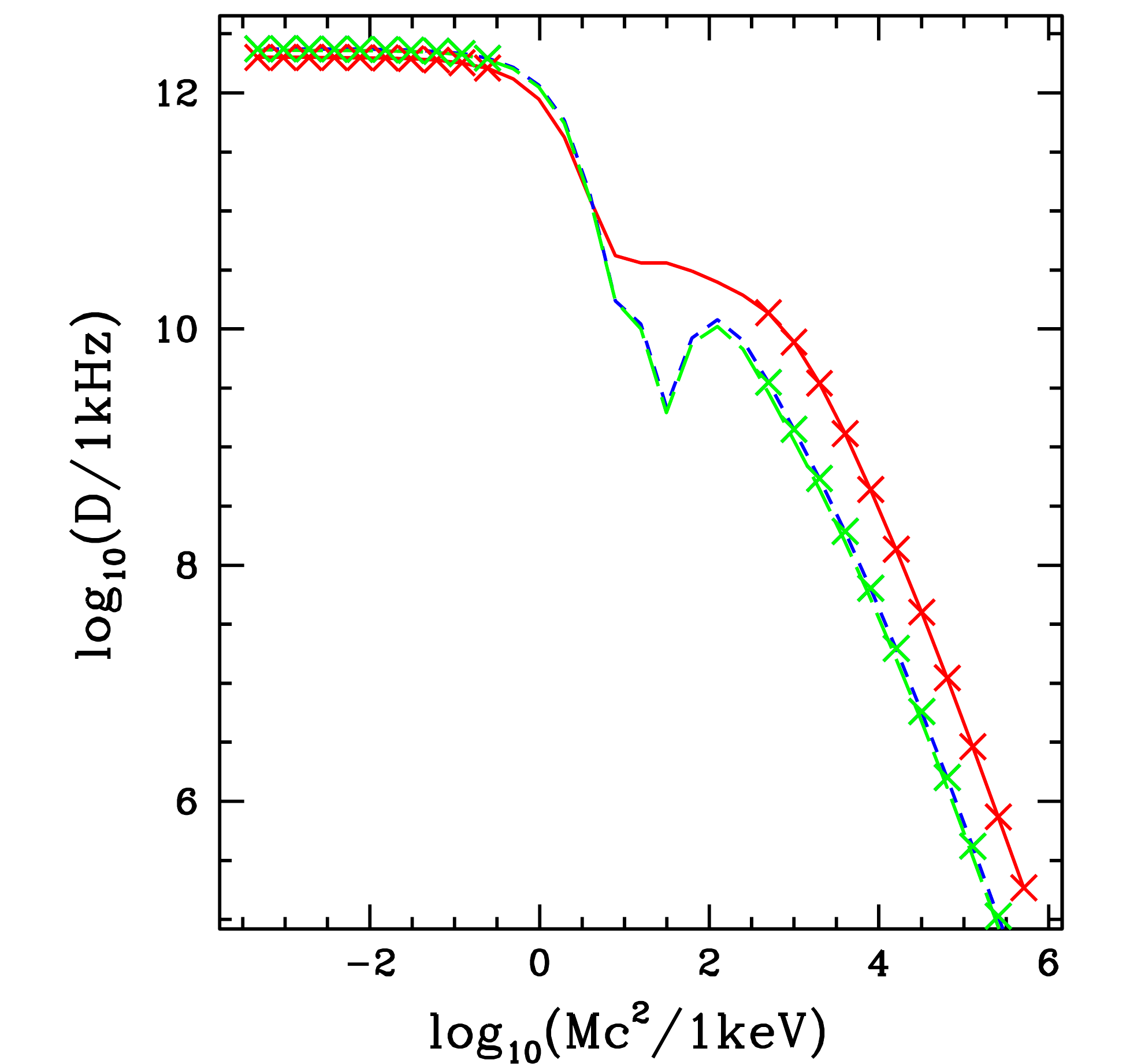}
	\caption{Visualization of the data from Table~\ref{t:dybii} for the electronic structure factors $D$ related to new boson. Solid and dashed lines correspond to the results of numerical calculations. Crosses represent the results of analytical estimations~\cite{FlambaumPRA2018}. Red colour corresponds to the $6s$ state, blue and green colours correspond to the $5d_{3/2}$ and $5d_{5/2}$ states respectively.}
	\label{f:dybii}
\end{figure}

We check our results for the factors $D$ against analytical estimations of Ref.~\cite{FlambaumPRA2018} for the limiting cases of small and large boson mass. This comparison is illustrated for the case of Yb$^+$ in Fig.~\ref{f:dybii} (see also Table~\ref{t:dybii}). Similar agreement between numerical and analytical results for the factors $D$ is achieved for the case of Yb.

The values $D_a$ and $D_b$ for the transitions in Yb$^+$ at small boson mass are also in good agreement with those of Ref.~\cite{AllehabiPRA2021} (the MBPT case, Table S4). Note that due to a difference in the definitions of the factors, for the sake of comparison, the values from Ref.~\cite{AllehabiPRA2021} should be multiplied by $\alpha\approx1/137$. 

\begin{table*}
	\caption{Electronic structure factors $D$ for the $6s$, $5d_{3/2}$, and $5d_{5/2}$ states of Yb$^+$ (kHz). Comparison of numerical calculations with analytical estimations from Ref.~\cite{FlambaumPRA2018}. For lines 1 to 5 the Yukawa potential is reduced to $\delta$-function and formulas (C1) and (C2) from Ref. \cite{FlambaumPRA2018} are used for electron density at the nucleus; formula (50) is used for lines 6 to 11, and formula (48) is used for lines 22 to 31.}
	\label{t:dybii}
	\begin{ruledtabular}
		\begin{tabular}{rl ll ll ll}
			\multicolumn{1}{c}{$N$} &
			\multicolumn{1}{c}{$M$} &
			\multicolumn{2}{c}{$D_{6s}$ [kHz]} &
			\multicolumn{2}{c}{$D_{5d_{3/2}}$ [kHz]} &
			\multicolumn{2}{c}{$D_{5d_{5/2}}$ [kHz]} \\
			&\multicolumn{1}{c}{keV} &
			\multicolumn{1}{c}{this work} &
			\multicolumn{1}{c}{\cite{FlambaumPRA2018}} &
			\multicolumn{1}{c}{this work} &
			\multicolumn{1}{c}{\cite{FlambaumPRA2018}} &
			\multicolumn{1}{c}{this work} &
			\multicolumn{1}{c}{\cite{FlambaumPRA2018}} \\
			\hline
			1 &  0.5110E+06 &  0.1858E+06 & 0.1989E+06 &  0.2667E+05 & 0.1636E+06 & 0.2194E+05 & 0.8079E+06 \\ 
			2 &  0.2555E+06 &  0.7395E+06 & 0.7955E+06 &  0.1061E+06 & 0.6543E+06 & 0.8732E+05 & 0.3231E+07 \\
			3 &  0.1277E+06 &  0.2906E+07 & 0.3182E+07 &  0.4171E+06 & 0.2617E+07 & 0.3432E+06 & 0.1293E+08 \\
			4 &  0.6387E+05 &  0.1107E+08 & 0.1273E+08 &  0.1589E+07 & 0.1047E+08 & 0.1308E+07 & 0.5170E+08 \\
			5 &  0.3194E+05 &  0.4003E+08 & 0.5091E+08 &  0.5755E+07 & 0.4187E+08 & 0.4732E+07 & 0.2068E+09 \\
			6 &  0.1597E+05 &  0.1363E+09 & 0.1531E+09 &  0.1967E+08 & 0.1259E+09 & 0.1615E+08 & 0.6220E+09 \\
			7 &  0.7984E+04 &  0.4370E+09 & 0.5042E+09 &  0.6352E+08 & 0.4147E+09 & 0.5205E+08 & 0.2048E+10 \\
			8 &  0.3992E+04 &  0.1302E+10 & 0.1660E+10 &  0.1929E+09 & 0.1366E+10 & 0.1574E+09 & 0.6745E+10 \\
			9 &  0.1996E+04 &  0.3470E+10 & 0.5468E+10 &  0.5428E+09 & 0.4497E+10 & 0.4395E+09 & 0.2221E+11 \\
			10 &  0.9980E+03 &  0.7768E+10 & 0.1800E+11 &  0.1414E+10 & 0.1481E+11 & 0.1137E+10 & 0.7314E+11 \\
			11 &  0.4990E+03 &  0.1372E+11 & 0.5929E+11 &  0.3536E+10 & 0.4876E+11 & 0.2881E+10 & 0.2408E+12 \\
			12 &  0.2495E+03 &  0.1944E+11 &         &  0.7982E+10 &         & 0.6753E+10 &   \\
			13 &  0.1248E+03 &  0.2486E+11 &         &  0.1195E+11 &         & 0.1053E+11 &   \\
			14 &  0.6238E+02 &  0.3106E+11 &         &  0.8351E+10 &         & 0.7570E+10 &  \\
			15 &  0.3119E+02 &  0.3623E+11 &         &  0.2188E+10 &         & 0.1961E+10 &  \\
			16 &  0.1559E+02 &  0.3639E+11 &         &  0.1097E+11 &         & 0.1000E+11 &  \\
			17 &  0.7797E+01 &  0.4183E+11 &         &  0.1722E+11 &         & 0.1729E+11 &  \\
			18 &  0.3899E+01 &  0.1264E+12 &         &  0.1401E+12 &         & 0.1281E+12 &  \\
			19 &  0.1949E+01 &  0.4225E+12 &         &  0.5860E+12 &         & 0.5549E+12 &  \\
			20 &  0.9746E+00 &  0.8829E+12 &         &  0.1161E+13 &         & 0.1116E+13 &  \\
			21 &  0.4873E+00 &  0.1315E+13 &         &  0.1646E+13 &         & 0.1594E+13 &  \\
			22 &  0.2437E+00 &  0.1616E+13 & 0.2944E+13 &  0.1965E+13 & 0.2256E+13 & 0.1911E+13 & 0.2215E+13  \\
			23 &  0.1218E+00 &  0.1796E+13 & 0.2944E+13 &  0.2150E+13 & 0.2256E+13 & 0.2095E+13 & 0.2215E+13 \\
			24 &  0.6092E-01 &  0.1894E+13 & 0.2944E+13 &  0.2249E+13 & 0.2256E+13 & 0.2194E+13 & 0.2215E+13 \\
			25 &  0.3046E-01 &  0.1945E+13 & 0.2944E+13 &  0.2301E+13 & 0.2256E+13 & 0.2246E+13 & 0.2215E+13 \\
			26 &  0.1523E-01 &  0.1971E+13 & 0.2944E+13 &  0.2327E+13 & 0.2256E+13 & 0.2272E+13 & 0.2215E+13 \\
			27 &  0.7614E-02 &  0.1984E+13 & 0.2944E+13 &  0.2340E+13 & 0.2256E+13 & 0.2285E+13 & 0.2215E+13 \\
			28 &  0.3807E-02 &  0.1991E+13 & 0.2944E+13 &  0.2347E+13 & 0.2256E+13 & 0.2292E+13 & 0.2215E+13 \\
			29 &  0.1904E-02 &  0.1994E+13 & 0.2944E+13 &  0.2350E+13 & 0.2256E+13 & 0.2295E+13 & 0.2215E+13 \\
			30 &  0.9518E-03 &  0.1997E+13 & 0.2944E+13 &  0.2352E+13 & 0.2256E+13 & 0.2297E+13 & 0.2215E+13 \\
			31 &  0.4759E-03 &  0.1999E+13 & 0.2944E+13 &  0.2356E+13 & 0.2256E+13 & 0.2301E+13 & 0.2215E+13 \\
		\end{tabular}
	\end{ruledtabular}
\end{table*}

\begin{table}
	\caption{Calculated electronic structure factors $D_a$ ($6s-5d_{5/2}$ transition in Yb$^+$), $D_b$ ($6s-5d_{3/2}$ transition in Yb$^+$), and $D_c$ ($^1$S$_0 - ^1$D$_2$ transition in Yb).}
	\label{t:npv}
	\begin{ruledtabular}
		\begin{tabular}{cccc}
			\multicolumn{1}{c}{$M$ (keV)} &
			\multicolumn{1}{c}{$D_b$ (kHz)} &
			\multicolumn{1}{c}{$D_a$ (kHz)} &
			\multicolumn{1}{c}{$D_c$ (kHz)} \\
			\hline
			0.3613E+06 &    0.4242E+06 &  0.4148E+06 &  0.3429E+06 \\
			0.1807E+06 &    0.1681E+07 &  0.1643E+07 &  0.1358E+07 \\
			0.9034E+05 &    0.6521E+07 &  0.6376E+07 &  0.5270E+07 \\
			0.4517E+05 &    0.2425E+08 &  0.2371E+08 &  0.1960E+08 \\
			0.2258E+05 &    0.8513E+08 &  0.8322E+08 &  0.6879E+08 \\
			0.1129E+05 &    0.2816E+09 &  0.2752E+09 &  0.2275E+09 \\
			0.5646E+04 &    0.8739E+09 &  0.8536E+09 &  0.7054E+09 \\
			0.2823E+04 &    0.2489E+10 &  0.2428E+10 &  0.2005E+10 \\
			0.1412E+04 &    0.6213E+10 &  0.6042E+10 &  0.4982E+10 \\
			0.7059E+03 &    0.1289E+11 &  0.1246E+11 &  0.1021E+11 \\
			0.3529E+03 &    0.2214E+11 &  0.2120E+11 &  0.1709E+11 \\
			0.1765E+03 &    0.3262E+11 &  0.3119E+11 &  0.2465E+11 \\
			0.8824E+02 &    0.3910E+11 &  0.3792E+11 &  0.3027E+11 \\
			0.4412E+02 &    0.3869E+11 &  0.3829E+11 &  0.3212E+11 \\
			0.2206E+02 &    0.4059E+11 &  0.4017E+11 &  0.3438E+11 \\
			0.1103E+02 &    0.5704E+11 &  0.5579E+11 &  0.4499E+11 \\
			0.5515E+01 &    0.3786E+11 &  0.4223E+11 &  0.4262E+11 \\
			0.2758E+01 &   -0.8656E+11 & -0.6509E+11 &  0.1386E+10 \\
			0.1379E+01 &   -0.2295E+12 & -0.1902E+12 & -0.3169E+11 \\
			0.6894E+00 &   -0.3109E+12 & -0.2615E+12 & -0.3833E+11 \\
			0.3447E+00 &   -0.3424E+12 & -0.2891E+12 & -0.3637E+11 \\
			0.1724E+00 &   -0.3523E+12 & -0.2977E+12 & -0.3472E+11 \\
			0.8618E-01 &   -0.3551E+12 & -0.3001E+12 & -0.3407E+11 \\
			0.4309E-01 &   -0.3558E+12 & -0.3008E+12 & -0.3387E+11 \\
			0.2155E-01 &   -0.3560E+12 & -0.3009E+12 & -0.3381E+11 \\
			0.1077E-01 &   -0.3560E+12 & -0.3010E+12 & -0.3380E+11 \\
			0.5387E-02 &   -0.3561E+12 & -0.3010E+12 & -0.3380E+11 \\
			0.2694E-02 &   -0.3561E+12 & -0.3010E+12 & -0.3381E+11 \\
			0.1347E-02 &   -0.3564E+12 & -0.3012E+12 & -0.3396E+11 \\
			0.6734E-03 &   -0.3556E+12 & -0.3005E+12 & -0.3338E+11 \\
			0.3367E-03 &   -0.3573E+12 & -0.3020E+12 & -0.3397E+11 \\
		\end{tabular}
	\end{ruledtabular}
\end{table}

\subsection{Sources of King plot nonlinearity}

In Fig.~3 of the main text we graph the nonlinearity measures $\zeta_+$ against $\zeta_-$. In this graph we also show straight lines passing through the origin that represent the directions that would result from various different sources of NL. We can do this because the electronic factors are the same for each pair of isotopes, and so they factor out of the nonlinearity measures if only one nonlinearity term in~(3) dominates. The ratio $\zeta_-/\zeta_+$ will be the same for all frequency-normalised transitions using the same reference transition (we use transition $a$, Yb$^+$ $6s \rightarrow 5d_{5/2}$,  as the reference).

We see from Fig.~3 that the ratio $\zeta_-/\zeta_+ = -1.5\,(4)$ given by our measurements does not agree with the ratios given by an additional light boson or by quadratic field shift (a combination of the two is still a possibility). On the other hand we see that the ratio suggested by nuclear deformation lies quite close to it. To determine these ratios we require values of the nuclear parameters  $\langle r^2 \rangle$ and $\langle r^4 \rangle$ for each isotope of interest. We use experimental values of $\langle r^2 \rangle$ from \cite{Angeli2013} and values of nuclear deformation parameters $\beta$ from \cite{raman01adndt}. We fit $\langle r^4 \rangle$ by the formula~\cite{AllehabiPRA2021}
\begin{equation}
\label{e:b}
\langle r^4 \rangle =\left[b_0 + b_1(r_c^2-r_0^2) + b_2(\beta-\beta_0)  \right] r_c^4, 
\end{equation}
where $r_0 = 5.179~{\rm fm}$ and  $\beta_0=0.305$, $b_0=1.3129(1)$, $b_1=-0.0036(1)$, and $b_2=0.10(1)$, $r_c=\sqrt{\langle r^2 \rangle}$.
Parameters $r_0$ and $\beta_0$ are fixed, parameters $b_0$, $b_1$, $b_2$ are found from the fitting of the results of numerical integration of the nuclear density for a set of nuclear parameters $r_c$ and $\beta$ which varies within the range determined by nuclear theory. The uncertainties in $b_0$, $b_1$, $b_2$ reflect the accuracy of the fitting.

With these experimental values we find that $\overline{\delta\langle r^4 \rangle}$ has nonlinearity ratio of $\zeta_-/\zeta_+ = -1.2 (1.0)$, consistent with our data. Alternatively, using the FIT values determined in~\cite{AllehabiPRA2021} to fit the Yb$^+$ data, we obtain $-0.33$ which is in tension with our data.

Finally we note that the recent preprint~\cite{ono21arxiv}, submitted shortly before finalisation of this manuscript, presents isotope shift data for the $6s^2\ ^1\textrm{S}_0 \rightarrow 6s6p\ ^3\textrm{P}_0$ clock transition. These data give $\zeta_-/\zeta_+ = -1.73\,(4)$ for the rather large nonlinearity observed, which is also consistent with the nuclear deformation hypothesis.


\end{document}